\documentclass[12pt]{article}
\usepackage{authblk}
\usepackage{geometry}
\usepackage{amsmath}
\usepackage{amssymb}
\usepackage{amsthm}
\usepackage{mathrsfs}
\usepackage{mathtools}
\usepackage{subfig}
\usepackage{customcommands}
\usepackage{bm}
\usepackage{Nettastyle}

\preprint{MIT-CTP/5217}
\title{Free Energy from Replica Wormholes}
\author[1]{Netta Engelhardt,}
\author[2]{Sebastian Fischetti,}
\author[2]{and Alexander Maloney} 

\affiliation[1]{Center for Theoretical Physics,\\
Massachusetts Institute of Technology, Cambridge, MA 02139, USA}
\affiliation[2]{Department of Physics, McGill University, Montr\'eal, QC, H3A 2T8, Canada}

\emailAdd{engeln@mit.edu}
\emailAdd{fischetti@physics.mcgill.ca}
\emailAdd{maloney@physics.mcgill.ca}

\abstract{Euclidean wormholes -- geometries which connect disconnected boundaries -- present a challenge to a standard quantum mechanical interpretation of the theory. One potential resolution is that the gravitational path integral computes the ensemble average of many theories.  The connected topologies contribute to the simplest possible observable: the free energy, which is computed using a replica trick.  This is distinct from the replica trick used to compute entanglement entropies, and appears in the computation of any extensive quantity. We argue that both JT gravity and a simplified version of CGHS admit a regime where the contribution of connected replica wormholes to the free energy is larger than that of disconnected topologies.  In both theories we find evidence of replica symmetry breaking, which is reminiscent of the behavior of certain spin glasses.  We discuss possible insights about ensemble averaging in gravity from this perspective. 
}

\begin{document}

\maketitle

\section{Introduction}
\label{sec:intro}

The process by which information escapes from the black hole hole interior is a pivotal question in the study of quantum gravity.  Recent work has brought to light two important points on this front.  The first is that at least in some gravitational models the Euclidean gravitational path integral (GPI) exhibits traces of unitarity: a GPI calculation of the entropy of the Hawking radiation reproduces the unitary Page curve~\cite{Pen19, AEMM, BouTom19, PenShe19, AlmHar19, MarMax20, GidTur20, HarSha20, GauFri20}. This hinges crucially on the contribution from Euclidean replica wormhole saddles that connect disconnected boundaries.  The inclusion of such wormholes implies that absent some further UV effects, the GPI would not factorize across disconnected boundaries~\cite{Col88, GidStr88,GidStr88a, MalMao04, ArkOrg07}.  In this case, the GPI cannot be interpreted as computing the partition function of a standard quantum mechanical theory.  One possible explanation is that the GPI should instead be interpreted as computing the ensemble average of many different quantum theories (see also~\cite{HarJaf18,StaWit19, Ill19, KapMah19} for related discussions).  Another possibility is that additional contributions should be included, which would lead to the expected factorization of the Euclidean partition function on disconnected surfaces. 

Our goal in this paper is to understand more systematically how Euclidean wormholes influence the physics of the GPI.  We will put aside for the time being any further potential UV effects (such as certain doubly non-perturbative effects in JT gravity) which might be necessary to describe the dual of an individual quantum theory.  We will investigate the contribution of Euclidean wormholes to a more general -- and in a sense simpler -- class of observables than the entropies described above.  We find that, completely independently of any considerations of black hole physics, these wormholes make important (and apparently indispensable) contributions to the dynamics of the theory.

To understand how Euclidean wormholes contribute, let us imagine computing a Euclidean GPI where we sum over geometries with a particular choice of boundary~$B$:
\be
\Pcal(B) \equiv \int_{\partial M=B} Dg \, e^{-S}.
\ee
We will take $B$ to be a connected surface, so this is usually interpreted as giving the gravitational computation of a partition function $Z(B)$.  One can also consider the integral over geometries with boundary $B^m = B \cup \dots \cup B$:
\be
\Pcal(B^m) \equiv \int_{\partial M=B^m} Dg \, e^{-S}.
\ee
If Euclidean wormholes contribute, then $\Pcal(B^m) \ne \Pcal(B)^m$ and the resulting partition function (or, more generally, correlation functions) do not factorize. 
One potential interpretation is that the GPI computes ensemble averages:
\be
\label{eq:GPIaverage}
\Pcal(B) = \overline{Z(B)}, \qquad \Pcal(B^m) = \overline{Z(B)^m},
\ee
where the overline denotes the average over a family of unitary quantum theories, and $Z(B)$ the partition function of a member of this family\footnote{The details of the ensemble and how it is computed will depend on the gravitational theory.
In a specific case like JT gravity, we interpret~$Z(B)$ as~$\Tr e^{-\beta H}$ where~$H$ is a random Hermitian matrix over which we average to get~$\overline{Z(B)}$~\cite{SSS}; see~\cite{MalWit20,PerTro20,CotJen20} for somewhat similar examples in one higher dimension.}.  Here we will remain agnostic on whether this ensemble average is genuinely a feature of the gravitational theory, or whether it merely appears as an approximate contribution to the low-energy effective description of some UV-complete theory.  Nevertheless, we will continue to interpret the GPI as an ensemble average, bearing in mind that this interpretation may only be valid in some effective description.  We will revisit these issues in more detail in Section~\ref{sec:disc}.

Our first observation is that if Euclidean wormholes contribute to the GPI, then they should contribute to even the most basic observable of the theory: the free energy ~$F = - T \ln Z$ evaluated at a particular temperature $T$.  In particular, let us imagine computing the free energy via a GPI, where~$T$ enters through the choice of $B$ (for example, in a two-dimensional theory of gravity,~$B$ is a circle of length $\beta \equiv 1/T$).  Na\"ively, of course, one might try to compute it by simply taking
\be
F = F_\mathrm{ann} \equiv -T \ln \Pcal(B) = -T \ln \overline{Z(B)}.
\label{annealed}
\ee
This, however, is in tension with the ensemble interpretation: since~$\overline{Z(B)}$ involves an integration over the random variables defining a particular instance of the ensemble, we may interpret~$\overline{Z(B)}$ as the partition function of a theory in which the random variables themselves are permitted to fluctuate and come into equilibrium.  In condensed matter systems, the free energy~$F_\mathrm{ann}$ defined above is therefore interpreted as an \textit{annealed} free energy.  Instead, what one is really interested in is the \textit{quenched} free energy, in which the random variables defining a particular instance of the ensemble are not allowed to equilibrate.  In other words, the free energy~$F = -T \ln Z(B)$ is computed in a particular instance of the ensemble, and \textit{then} the average is taken:
\be
\overline{F} = -T \, \overline{\ln Z(B)}.
\label{quenched}
\ee
In general the annealed and quenched free energies will be different.  Indeed, from the gravitational point of view one might expect that $\overline{\ln Z(B)} \ne \ln {\overline{Z(B)}}$ whenever Euclidean wormholes are present in the theory, for the same reason that $\overline{Z}^m \ne \overline{Z^m}$.

In order to understand exactly how Euclidean wormholes contribute to (\ref{quenched}), one needs to compute~$\overline{F}$ from the GPI using a replica trick that involves considering the GPI on~$m$ copies of the boundary~$B$ and then analytically continuing to~$m = 0$.  This replica trick is distinct from the one that is employed to compute the von Neumann entropy (which instead considers the GPI defined by an~$n$-sheeted boundary manifold and then continues to near~$n = 1$), and a completely consistent calculation of entanglement entropy must implement \textit{both} replica tricks.  This version of the replica trick will be reviewed in section \ref{sec:replicatrick}, and is common in the condensed matter literature, especially in the study of spin glasses.
In fact, although we have focused on the free energy, this new replica trick will apply to the computation of any extensive observable.  For example, in the calculation of the Renyi entropy $S_{n}$ of a pure state from the GPI in \cite{AlmHar19}, the result vanishes only to leading order if this additional replica trick is not implemented: the Renyi entropy vanishes \textit{identically} only when the calculation correctly implements both replica tricks.

This additional replica trick means that~$\overline{F}$ becomes sensitive to the contribution of wormholes connecting the replicas, and leads to the conclusion that it is \textit{not consistent} to simultaneously interpret $\Pcal(B)$ as computing an ensemble average and to compute the free energy (or more generally, any extensive obervable) without including contributions from Euclidean wormholes.  If the free energy computation is dominated by the disconnected topology, then the ensemble averaging leaves no visible footprint, and the quenched free energy coincides with the annealed free energy:~$\overline{F} \approx -T \ln \Pcal(B)$.  However, if in some regime replica wormholes contribute nontrivially to~$\overline{F}$, then ensemble averaging is important for the computation of \textit{any} observable in that regime. Failure to properly compute the free energy via the replica trick above will erase subtle signatures of the ensemble.

Of course, the skeptical reader may be concerned that replica wormholes might \textit{never} actually make an appreciable contribution to~$\overline{F}$, at least in those regimes in which we have some control over the gravitational theory.  Indeed, although it has now been verified that replica wormholes are important in the study of black hole entropy, it need not follow that such wormholes will be important in the computation of~$\overline{F}$.

To address  this potential concern, in Sections~\ref{sec:CGHS} and~\ref{sec:JT} we compute the free energy in two different models of~2D gravity.  We find that the na\"ive calculation of the annealed free energy~$F_\mathrm{ann}$ exhibits pathological behavior at sufficiently low temperature.  Specifically, it is non-monotonic with temperature, implying a negative thermodynamic entropy~$S = - \partial F/\partial T$.  We then use the replica trick to investigate the contribution of replica wormholes to~$\overline{F}$, finding that this contribution becomes larger than that of the disconnected topology when the annealed free energy exhibits its unphysical behavior.  The inclusion of wormholes ameliorates the pathological behavior of the free energy at low temperature, at least with a certain implementation of the replica trick.

The gravitational systems that we consider are~$\widehat{\mathrm{CGHS}}$~\cite{CGHS, CGHShat} and JT gravity~\cite{Tei83, Jackiw}, and importantly we compute the free energy using the full GPI (computed for~$\widehat{\mathrm{CGHS}}$ in~\cite{GodMar20} and JT gravity in~\cite{SSS}), rather than a saddle-point approximation. In both models, we find that replica wormholes substantially change the behavior of the free energy at sufficiently low temperature.  Interestingly, in JT gravity, we find that the temperature at which the pathological behavior of the disconnected free energy manifests, and the temperature at which contributions from replica wormholes dominate, both scale like~$e^{-2S_0/3}$ (where~$e^{-S_0}$ controls the JT gravity genus expansion).  Since the gravitational theory is only under control for large~$S_0$, one might be concerned that the contribution of the replica wormholes happens in a regime of the theory in which we have no perturbative control.  In fact, working at large~$S_0$ but with~$T e^{2S_0/3}$ of order unity puts us in the so-called Airy limit, where the system is controlled by the universal behavior of the edge of the classical density of eigenvalues~$\rho_0(E)$\footnote{We will discuss subtleties involved in this limit in Section~\ref{subsec:Airy}.}.  In this limit, the genus expansion can be summed, providing a handle on doubly-nonperturbative corrections (in~$S_0$).  We find that these corrections are unimportant in part of the regime where replica wormholes dominate, so we can conclude that they genuinely do contribute even when doubly-nonperturbative corrections do not.  This story is entirely analogous to the replica wormholes narrative in the context of black hole evaporation: some parameter~$k$ parametrizing the entropy of matter must become nonperturbatively large in~$S_0$ in order for replica wormholes to dominate, and this transition happens right at the edge of validity of the semiclassical approximation.  In our context, the parameter that must become large for wormholes to dominate is instead the inverse temperature~$\beta$.

In an intriguing turn of events, while the replica wormholes do mitigate the pathologies in the free energy, we cannot show that they remove them entirely.  We argue that this is due to the inherent ambiguity in the analytic continuation that defines~$\overline{F}$.  To gain more insight into this ambiguity, in Section~\ref{sec:RSB} we point out that an extremely similar phenomenon happens in spin glass systems, where a quenched disorder can allow for the spontaneous coupling of replicas used to calculate~$\overline{F}$.  In that context, we  review the Sherrington-Kirkpatrick (SK) model of spin glasses, and note that similar to our gravity calculations, at high temperature the free energy is dominated by a paramagnetic phase in which the replicas are uncorrelated, while at sufficiently low temperatures the system enters a spin glass phase in which the replicas correlate\footnote{We should be quick to note that our gravitational results also exhibit some important qualitative differences from spin glasses, notably the fact that we need to go to nonperturbatively low temperature to see an exchange of dominance, while the spin glass phase transition happens at a temperature of order unity and can be seen in a strictly thermodynamic limit.}.  A replica-symmetric analysis of the spin glass phase exhibits the same sorts of pathologies that we see in the quenched free energy of~$\widehat{\mathrm{CGHS}}$ and JT gravity; it turns out that in the SK model, replica symmetry breaking (RSB) is the key structure that ``fixes'' the analytic continuation in the replica trick and gives the correct free energy down to zero temperature.  Motivated by the parallels between spin glasses and our gravitational results, we conjecture that the same sort of RSB is needed in the gravitational case to fully capture the correct behavior of~$\overline{F}$ at low temperature.  Importantly, the RSB that we discuss is notably different from the sort of RSB ordinarily discussed in the context of gravitational calculations of Renyi entropies.  We make more exploratory comments about possible parallels between gravity and spin glasses in Section~\ref{sec:disc}, but also note that our results should not necessarily be interpreted as indicative of a literal gravitational spin glass phase.

\paragraph{Relation to prior work:}  In the context of JT gravity, preludes of the transition in which we are interested can be found in analyses of the two-point correlator~$\overline{Z(\beta_1) Z(\beta_2)}$, which is relevant for studies of the spectral form factor.  For instance,~\cite{OkuSak19,OkuSak20} find that at temperatures lower than~$\Ocal(e^{-2S_0/3})$, the contribution of the cylinder topology to this correlator can become larger than that of the disk; see also~\cite{Joh20a,Joh20} for the same behavior in nonperturbative completions of JT gravity, without needing to work at large~$S_0$.  See also~\cite{Oku19} for an analogous transition in a Gaussian matrix model.  Our purpose here is specifically to investigate the contributions of connected topologies to the quenched free energy via the replica trick for~$\overline{\ln Z}$.

While we emphasize that we do not claim a bona fide spin glass phase in JT gravity, the behavior is sufficiently similar that further comment is warranted given recent studies on SYK.  These investigations show that SYK does not exhibit a spin glass phase; that is, a saddle-point analysis of the replica trick in the large-$N$ limit (see e.g.~\cite{BagAlt16,KitSuh17}) indicates that no saddles correlating different replicas dominate the correlators~$\overline{Z^m}$ at any temperature~\cite{MalSta16,GarVer16,BagAlt16,CotGur16,GurMah18,AreKhr18,CarCar18,Ye18,GeoPar01,FuSac16}.  Here we point out that (i) we do not work in a saddle-point approximation, and in fact we expect that the behavior we study would be invisible in such a limit; and (ii) JT gravity is only dual to a low-energy regime of SYK, and as shown in~\cite{AreKhr18} an appropriate IR limit of SYK can exhibit a different phase structure than the full SYK system.  Hence there is no tension with our results.

More generally, attempts to model spin glasses holographically, such as e.g.~\cite{FujHik08, AhaKom15}, typically manually turn on a correlation between the different replica boundaries in order to induce a spin glass phase transition; this is analogous to the correlation between replica boundaries that occurs in computations of the entropy of Hawking radiation (due to tracing out a subsystem), or to the coupling of two boundaries in the traversable wormhole setup of~\cite{GaoJaf16, MalQi18}.  Here we are specifically interested in the contribution of replica wormholes to the GPI~$\Pcal(B^m)$ defined by~$m$ \textit{completely uncoupled} boundaries: the coupling happens entirely spontaneously and is an inevitable consequence of replica wormholes.
 
On a more tangential note, let us finally point out that there has been an ongoing discussion of the relevance of spin glasses to the physics of eternal inflation as well as to the landscape of string vacua. See e.g.~\cite{AnnDen11} as well as ~\cite{DenTASI} for an excellent review, and also ~\cite{JaiVan15} for more recent work. In a similar vein,~\cite{AnnAno13b} discussed these topics in the context of AdS$_{2}$, and \cite{AnnAno11} and \cite{AnnAno13a} studied a spin glass phase of black hole microstates (without external coupling).  It would be interesting to explore connections to our present work.

\section{The Replica Trick for $\overline{\ln Z}$}
\label{sec:replicatrick}

The purpose of this section is to discuss in more detail the replica trick necessary for the computation of the free energy~$\overline{F}$, and more generally the ensemble average of the generating functional~$\overline{\ln Z}$ considered as an arbitrary function of sources.  Since such an average appears in the computation of Renyi entropies~$S_n$, and hence also of the von Neumann entropy, we will also discuss the relation to the replica trick used in the computation of von Neumann entropy.

The key point is that if the GPI is interpreted as the ensemble average of a partition function as per~\eqref{eq:GPIaverage}, then it cannot directly compute the ensemble average of any extensive quantity, such as~$\overline{\ln Z}$.  The replica trick relates such extensive observables to non-extensive objects via
\be
\label{eq:replicatrick}
\overline{\ln Z(B)} = \lim_{m \to 0} \frac{1}{m} \left(\overline{Z(B)^m}-1\right)=\lim\limits_{m\rightarrow 0} \frac{1}{m} (\Pcal(B^{m})-1),
\ee
where~$B^m$ denotes~$m$ copies of the boundary~$B$, and we have assumed that the pre-average partition function obeys~$Z(B)^m = Z(B^m)$; that is, that~$m$ copies of the (non-averaged) partition function on the boundary~$B$ can equivalently be expressed as the partition function of~$m$ copies of~$B$ (this is certainly the case if~$Z(B)$ is the partition function of an ordinary QFT living on~$B$).

The implementation of this replica trick clearly yields different behaviors of~$\overline{\ln Z(B)}$ depending on whether connected topologies contribute nontrivially to~$\Pcal(B^m)$.  In general, we have
\be
\Pcal(B^m) = \Pcal(B)^m + \sum_{\substack{\mathrm{connected} \\ \mathrm{topologies}}},
\ee
where the first term comes from summing over geometries that leave all the replica copies of~$B$ disconnected from one another, while the sum represents integrals over geometries that connect two or more copies of the boundary (i.e.~replica wormholes)\footnote{It is sometimes suggested that the factorization problem of the GPI can be avoided if either the sum over connected topologies is supposed to be excluded, or if somehow it conspires to give a vanishing contribution to~$\Pcal(B^m)$.  Here we adopt the perspective of~\cite{MarMax20} that excluding the connected topologies requires a non-local constraint, while having their collective contribution vanish would require fine-tuning.}.  We therefore generically have~$\Pcal(B^m) \neq \Pcal(B)^m$.  However, in certain cases one topological sector may dominate over others.  If the dominant contribution is disconnected, then 
we have
\be
\Pcal(B^m) \approx \Pcal(B)^m.
\ee
In this case, using~\eqref{eq:replicatrick} we see that~$\overline{\ln Z} \approx \ln \overline{Z}$, so the replica trick has no appreciable effect; in the condensed matter language used in Section~\ref{sec:intro}, the quenched free energy and the annealed free energy approximately coincide.  In particular, we may compute the gravitational free energy by just just taking~$\overline{F} \approx -T \ln \Pcal(B)$, as usual.  On the other hand, if a topology connecting multiple copies of~$B$ dominates, then we should expect that
\be
\Pcal(B^m) \not\approx \Pcal(B)^m,
\ee
so the quenched and annealed free energies should not even approximately coincide, and a proper computation of the gravitational free energy will not coincide with the annealed free energy:~$\overline{F} \not\approx -T \ln \Pcal(B)$.

\begin{figure}
\centering
\includegraphics[page=1,width=0.8\textwidth]{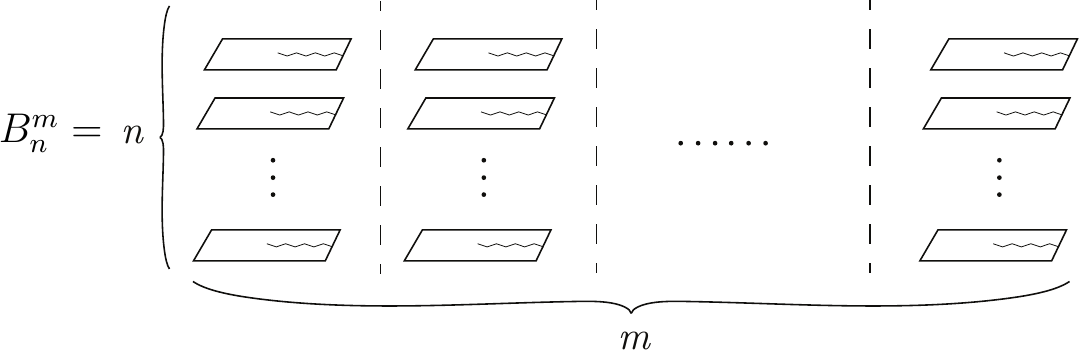}
\caption{A computation of the Renyi entropy~\eqref{eq:Renyiaverage} from the GPI requires an additional replica trick, involving computing the GPI with the boundary~$B_n^m$ shown here.  Each of the columns is an~$n$-sheeted geometry~$B_n$ constructed by slicing~$n$ copies of~$B$ along the region~$R$ and then identifying these copies cyclically along the cut.  $B_n^m$ consists of~$m$ copies of this multi-sheeted geometry.  The disorder-averaged von Neumann entropy is computed in the double limit~$m \to 0$,~$n \to 1$.}
\label{fig:ManyReplicas}
\end{figure}

Let us now exhibit how the replica trick~\eqref{eq:replicatrick} relates to the one used to compute the von Neumann entropy.  This latter replica trick defines the von Neumann entropy of a subsystem (say a region~$R \subset B$) as a limit of Renyi entropies:
\be
S = \lim_{n \to 1} S_n,
\ee
where the Renyi entropies~$S_n$ are given by
\be
S_n \equiv \frac{1}{1-n} \left(\ln Z(B_n) - n \ln Z(B) \right),
\ee
with~$B_n$ an~$n$-sheeted geometry consisting of~$n$ copies of~$B$ cut along the region~$R$ and then cyclically identified along this cut; see Figure~\ref{fig:ManyReplicas}.  If~$R$ is empty,~$B_n$ is just~$B^n$, consisting of~$n$ copies of~$B$.

Suppose we now wish to evaluate the Renyi entropies via a gravitational path integral, under the interpretation that it computes an \textit{ensemble average} of~$S_n$ (and hence also of the von Neumann entropy).  Such a computation requires the ensemble averages~$\overline{\ln Z(B_n)}$ and~$\overline{\ln Z(B)}$, which in turn requires use of the ``extra'' replica trick~\eqref{eq:replicatrick}:
\be
\label{eq:Renyiaverage}
\overline{S_n} = \frac{1}{1-n} \left(\lim_{m \to 0} \frac{1}{m} \left(\Pcal(B_n^m) - 1\right) - n \lim_{m \to 0} \frac{1}{m} \left(\Pcal(B^m) - 1\right)\right),
\ee
where~$B_n^m$ consists of~$m$ separate copies of the~$n$-sheeted geometry~$B_n$, as shown in Figure~\ref{fig:ManyReplicas}.  A correct calculation of the von Neumann entropy therefore \textit{requires} taking the double limit~$m \to 0$,~$n \to 1$.

A key distinction to note here is that the~$m$ replicated boundaries~$B_n^m$ are \textit{completely disconnected}; any geometric connection between them must come spontaneously from the GPI.  On the other hand, when~$R$ is non-empty, the geometry~$B_n$ is a single connected geometry, due to the identification of the~$n$ sheets along the cut~$R$.  In fact, when~$R$ is the empty set, the Renyi entropies must vanish exactly (since we are computing the entropy of a pure state); it is precisely the auxiliary replica trick over~$m$ that guarantees this.  To see this, note that if~$R$ is empty,~$B_n^m = B^{nm}$, and hence
\be
\lim_{m \to 0} \frac{1}{m} \left(\Pcal(B_n^m) - 1\right) = \left.\frac{\partial \Pcal(B^{nm})}{\partial m} \right|_{m = 0} = n \left. \frac{\partial \Pcal(B^{\widetilde{m}})}{\partial \widetilde{m}} \right|_{\widetilde{m} = 0},
\ee
so the two terms in~\eqref{eq:Renyiaverage} cancel identically, giving~$\overline{S_n} = 0$ for all~$n$.  Importantly, the vanishing of the Renyi entropy is independent of the dominant topology contributing to the path integral.  This should be contrasted with, for example, the computation of Renyi entropy performed in~\cite{AlmHar19}, which (working in a semiclassical regime) claimed that the entropy a pure state vanishes because in that case the GPI is dominated only by disconnected topologies.  The trouble with that interpretation is that even when the disconnected topology dominates, the path integral will still receive subdominant corrections from connected topologies which would lead to a nonvanishing (but small) Renyi entropy.  The double replica trick makes clear that the Renyi entropy of a pure state vanishes exactly, and even when the dominant geometry is a replica wormhole.

Of course, the claim of~\cite{AlmHar19} that (at least in their JT gravity model) the disconnected topology dominates the gravitational path integral in a semiclassical limit when~$R$ is the empty set might lead to a concern: even if replica wormholes make subdominant contributions to the free energy, they might never be dominant in a regime in which the gravitational theory is under control.  If so, the extra replica trick~\eqref{eq:replicatrick} will in practice never be necessary for computing leading-order effects.  To address this concern, we will now explore explicit examples of gravitational models in which connected saddles do make dominant contributions when the theory is at least somewhat under control, focusing specifically on computations of the free energy~$\overline{F}$.

\section{Free Energy in $\widehat{\mathrm{CGHS}}$}
\label{sec:CGHS}

\begin{figure}[t]
\centering
\subfloat[][]{
\includegraphics[width=0.2\textwidth,page=2]{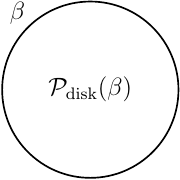}
\label{subfig:CGHSdisk}
}%
\hspace{3cm}
\subfloat[][]{
\includegraphics[width=0.25\textwidth,page=3]{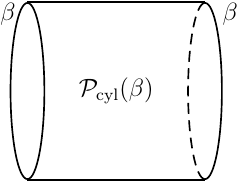}
\label{subfig:CGHScyl}
}
\caption{The only topologies that can appear in the~$\widehat{\mathrm{CGHS}}$ path integral are the disk and the cylinder.}
\label{fig:CGHS}
\end{figure}

We begin the investigation in gravity with a variant of standard CGHS dilaton gravity~\cite{CGHS}, introduced as the $\widehat{\mathrm{CGHS}}$ model in~\cite{AfsGon19} (following~\cite{CanJac92}).  This model is given by the Euclidean action
\be
S = \frac{\kappa}{2} \int d^2 x \, \sqrt{g} \left(\Phi R - 2\Psi + 2 \Psi \varepsilon^{\mu\nu} \partial_\mu A_\nu\right) + S_\partial,
\ee
where~$S_\partial$ is a boundary term.  The equation of motion for~$A_\mu$ fixes~$\Psi$ to be constant, and it is this constant value that sets the temperature of black hole solutions, while the equation of motion for~$\Phi$ sets~$R = 0$.   In fact, even in the path integral the integration over~$\Phi$ means that only strictly flat geometries contribute.  Hence the only contributions can come from the disk or the cylindrical topology, corresponding to one and two boundaries, respectively; see Figure~\ref{fig:CGHS}.  It is this simplification that will allow us to make definitive statements about the structure of the replicas and free energy in this model, without needing to worry about nonperturbative effects arising from higher-genus contributions.  This section is therefore a warmup for the JT gravity calculation in Section~\ref{sec:JT}, which is complicated by contributions from all topologies.

\subsection{Path integrals in $\widehat{\mathrm{CGHS}}$}
\label{subsec:CGHSpathintegral}

The path integrals of the disk and cylinder in~$\widehat{\mathrm{CGHS}}$ were computed in~\cite{GodMar20}.  
For the disk with boundary length~$\beta$, the result is\footnote{In the notation of \cite{GodMar20} we have chosen units where the coupling $\gamma$ (which is related to the boundary value of the dilaton) has been set equal to one, and where the normalization factor $\alpha$ which appears in the symplectic form is also equal to $1$.}
\be
\Pcal_\mathrm{disk}(\beta) = \frac{2\pi}{\beta^2}.
\ee
Already we can deduce the need for a phase transition.  If the disk were to dominate the free energy, we would have~$\overline{F} = -T \ln \Pcal_\mathrm{disk}$, which is clearly a non-monotonic function of temperature: it has a local maximum at~$T_\mathrm{max} = 1/\sqrt{2\pi} e$, corresponding to a negative thermodynamic entropy~$-\partial F/\partial T$ when~$T < T_\mathrm{max}$ (in fact, the entropy is logarithmically divergent at~$T = 0$).  We might hope that the contribution of the cylinder will rectify this low-temperature behavior.

To that end, the path integral on the cylinder (each of whose boundaries has length~$\beta$) is
\be
\Pcal_\mathrm{cyl}(\beta) = \frac{2\pi^2}{\beta}.
\ee
Let us use~$\Pcal_m(\beta)$ to denote the GPI defined by~$m$ boundaries of length~$\beta$.  This path integral receives competing contributions from the disk and the cylinder; the completely disconnected topology gives a contribution of
\be
\Pcal_m(\beta) \supset \Pcal_\mathrm{disk}(\beta)^m = \left(\frac{2\pi}{\beta^2}\right)^m,
\ee
while the topology that connects~$m/2$ pairs of boundaries with cylinders (temporarily taking~$m$ to be even) gives a contribution
\be
\Pcal_m(\beta) \supset \Pcal_\mathrm{cyl}(\beta)^{m/2} = \left(\frac{2\pi^2}{\beta} \right)^{m/2}.
\ee
At temperatures larger than~$T_c \equiv 2^{-1/3}$, the contribution from the disk topology is larger, while for temperatures smaller than~$T_c$, the contributions from the cylinder topology is larger.  So already at the level of this rough analysis we see a transition: the high-temperature behavior is controlled by the disconnected topology, while the low-temperature behavior is controlled by a connected one\footnote{Because this computation is done using the full path integral, there is no sense in which we can interpret these as saddles, with one ``dominating'' over the other.  The point is that both topologies contribute nontrivially, and for sufficiently large or small temperatures one contributes substantially more than the other.  The transition between these two behaviors cannot be expected to be sharp, of course.}.  Importantly,~$T_c > T_\mathrm{max}$, so the contribution from the cylinder modifies the free energy in the temperature regime in which the annealed free energy~$F_\mathrm{ann} \equiv -T\ln \Pcal_\mathrm{disk}$ was pathological.

Now let us be more thorough and compute~$\Pcal_m(\beta)$ exactly, therefore attempting to obtain the free energy via the~$m \to 0$ limit~\eqref{eq:replicatrick}.  Defining~$r \equiv \Pcal_\mathrm{cyl}/\Pcal_\mathrm{disk}^2$, we have
\be
\label{eq:PCGHSsum}
\Pcal_m(\beta) = \Pcal_\mathrm{disk}^m \sum_{m' = 0}^{\lfloor m/2 \rfloor} \begin{pmatrix} m \\ 2m' \end{pmatrix} (2m'-1)!! \, r^{m'},
\ee
where the sum counts contributions from all aways of connecting an even number~$2m'$ of boundaries together via cylinders, the binomial coefficient counts the ways of choosing~$2m'$ boundaries from the full set of~$m$, and the double factorial counts how many distinct ways there are of connecting those~$2m'$ boundaries pairwise with cylinder topologies.  Expressing the double factorial as
\be
(2m'-1)!! = \frac{2^{m'}}{\sqrt{\pi}} \, \Gamma\left(m'+\frac{1}{2}\right) = \int_0^\infty \frac{dt}{\sqrt{\pi t}} \, (2t)^{m'} e^{-t},
\ee
we find
\be
\Pcal_m(\beta) = \Pcal_\mathrm{disk}^m \int_0^\infty \frac{dt}{\sqrt{\pi t}} \, e^{-t} \sum_{m' = 0}^{\lfloor m/2 \rfloor} \begin{pmatrix} m \\ 2m' \end{pmatrix} (2tr)^{m'}.
\ee
The sum can be evaluated using the identity\footnote{\eqref{eq:identity} can be shown by expanding the binomials on the right-hand side and then using the identity for sums of roots of unity:
\be
\sum_{j = 0}^{M-1} \left(e^{2\pi j i/M}\right)^k = \begin{cases} 0, & k \in \mathbb{Z} \text{ and } k \neq 0 \text{ (mod }M) \\
M, & k \in \mathbb{Z} \text{ and } k = 0 \text{ (mod }M) \end{cases}. \nonumber
\ee}
\be
\label{eq:identity}
\sum_{m' = 0}^{\lfloor m/M \rfloor} \begin{pmatrix} m \\ M m' \end{pmatrix} y^{M m'} = \frac{1}{M} \sum_{j = 0}^{M-1} \left(1 + e^{2j\pi i/M} y\right)^m
\ee
for any positive integers~$m$ and~$M$, resulting in
\be
\label{eq:PbetamCGHS}
\Pcal_m(\beta) = \Pcal_\mathrm{disk}^m \int_0^\infty \frac{dt}{2\sqrt{\pi t}} \, e^{-t} \left(\left(1+\sqrt{2tr}\right)^m + \left(1-\sqrt{2tr}\right)^m\right).
\ee
To compute $\overline{\ln Z}$, we want to now continue to $m\rightarrow 0$.

\subsection{Continuing to non-integer $m$}
\label{subsec:CGHScontinuation}

The result~\eqref{eq:PbetamCGHS} can be naturally continued to non-integer~$m$, but it exhibits a curious feature: because the second term~$1-\sqrt{2tr}$ will always become negative somewhere in the region of integration, for non-integer~$m$ this term need not be (and is not) real.  Invoking the replica trick~\eqref{eq:replicatrick} at this stage would then yield a complex free energy, which is manifestly unphysical.  Evidently, the obvious analytic continuation of~\eqref{eq:PbetamCGHS} to non-integer~$m$ cannot be the correct one for the replica trick.  A more well-behaved alternative can be obtained by noting the following.  For any analytic function~$f(z)$ of a complex variable~$z$, let~$f^*(z)$ be the function obtained by complex-conjugating the Taylor series coefficients of~$f(z)$; then by construction the function~$f_r(z) \equiv (f(z) + f^*(z))/2$ is also analytic, and is real whenever~$z$ is.  If~$f(z)$ is real when~$z$ is a positive integer, then~$f_r(z) = f(z)$ when~$z$ is a positive integer, and both~$f(z)$ and~$f_r(z)$ therefore give admissible analytic continuations from the positive integers to general complex~$z$.  For this reason, for the purposes of computing~$\overline{F}$ via the replica trick we are free to simply use the real part of~\eqref{eq:PbetamCGHS} when~$m$ is real, which gives
\begin{multline}
\label{eq:PbetamCGHSreal}
\Pcal_m(\beta) = \Pcal_\mathrm{disk}^m \left\{\int_0^\infty \frac{dt}{2\sqrt{\pi t}} \, e^{-t} \left(\left|1+\sqrt{2tr}\right|^m + \left|1-\sqrt{2tr}\right|^m\right) \right. \\ \left. - 2 \sin^2\left(\frac{\pi m}{2}\right) \int_{1/2r}^\infty \frac{dt}{2\sqrt{\pi t}} \, e^{-t} \left|1-\sqrt{2tr}\right|^m \right\}.
\end{multline}

It may seem that we have pushed the replica trick to a breaking point.  Of course there was always an infinite amount of freedom in how to continue the path integral~$\Pcal_m(\beta)$ from positive integer~$m$ to non-integer~$m$ near zero, but the implied hope was that a ``natural'' analytic continuation should present itself, and that this continuation should be the correct one for getting the physically correct free energy.  But the natural continuation of~\eqref{eq:PbetamCGHS} gives a complex free energy, and we had to introduce a rather ad hoc procedure for modifying the continuation to obtain~\eqref{eq:PbetamCGHSreal}.  What prevents us from, say, adding~$g(T) \sin(\pi m)$ to~$\Pcal_m(\beta)$ with~$g(T)$ an arbitrary function of temperature, and therefore getting whatever free energy we want?

This discomfort is well-justified, for there is an even more serious problem with the continuation of either~\eqref{eq:PbetamCGHS} or~\eqref{eq:PbetamCGHSreal} to general \textit{complex}~$m$.  In order to consistently interpret~$\Pcal_m(\beta)$ as giving the disorder average~$\overline{Z^m}$ of some power of the partition function, its behavior for purely imaginary~$m = i\alpha$ must be bounded since
\be
\left|\Pcal_{i\alpha} (\beta)\right| = \left| \overline{Z^{i\alpha}} \right| \leq \overline{\left|Z^{i\alpha} \right|} = 1,
\ee
where we have assumed that the disorder average is defined by a proper probability distribution (i.e.~one that is positive and normalized).  But while the terms on the first line of~\eqref{eq:PbetamCGHSreal} are bounded when~$m$ is imaginary, the term on the second line is not, and indeed it grows arbitrarily large for large imaginary~$m$.  So~\eqref{eq:PbetamCGHSreal} cannot be interpreted as the analytic continuation to complex~$m$ of an ensemble average~$\overline{Z^m}$ with respect to a positive and normalized probability distribution.

In principle we should therefore look for a different analytic continuation that is well-behaved for imaginary~$m$ and hope that, say, Carlson's theorem is sufficient to ensure uniqueness of this continuation\footnote{Carlson's theorem says that if a function~$f(z)$ is analytic in the right half-plane~$\mathrm{Re}(z) > 0$, grows more slowly than~$\sin(\pi z)$ on the imaginary axis and no faster than exponentially elsewhere in the right half-plane, and vanishes on the non-negative integers, then~$f(z)$ vanishes identically.}.  However, the growth of~\eqref{eq:PbetamCGHSreal} at large \textit{real}~$m$ excludes this possibility.  To see why, note that~\eqref{eq:PbetamCGHSreal} grows faster than exponentially in~$m$ at large real integer~$m$, which can be seen easily by, say, keeping only the~$m' = \lfloor m/2 \rfloor$ term in the sum~\eqref{eq:PCGHSsum}.  To try to prove that the analytic continuation to non-integer~$m$ must be unique (once we impose boundedness for imaginary~$m$), suppose we had two different analytic continuations~$\Pcal_m^{(1)}$ and~$\Pcal_m^{(2)}$, and let us try to show that their difference~$\Delta \Pcal_m$ must vanish.  This difference of course vanishes on the positive integers, and must also be bounded on the imaginary axis if both~$\Pcal_m^{(1)}$ and~$\Pcal_m^{(2)}$ are.  To invoke Carlson's theorem to conclude that~$\Delta \Pcal_m$ must vanish identically, we therefore only need to guarantee that~$\Delta \Pcal_m$ grows no faster than exponentially in the right half-plane; but this is not a condition we can enforce via any constraint on~$\Pcal_m^{(1)}$ and~$\Pcal_m^{(2)}$ due to their superexponential growth for integer~$m$, and hence Carlson's theorem cannot be invoked.

The ambiguity in finding the ``correct'' analytic continuation is a substantial obstacle that we will address in much more detail in Section~\ref{sec:RSB}; it will be interpreted as a signature of replica symmetry breaking.  For the time being, we will forge ahead by just using~\eqref{eq:PbetamCGHSreal}, assuming that the temperatures at which the quenched free energy is sensitive to contributions from the cylinder coincide with the temperatures at which the~$\Pcal_m(\beta)$, and therefore the free energy obtained from~\eqref{eq:PbetamCGHSreal}, are.  In proceeding in this way, we will be unable to determine what the correct form of the quenched free energy~$\overline{F}$ actually should be, but we can still investigate when contributions from the cylinder cause the quenched and annealed free energies to differ.

With this important caveat in mind, the free energy obtained from~\eqref{eq:PbetamCGHSreal} is
\be
\label{eq:CGHSfreenergy}
\overline{F} = -T\left(\ln \Pcal_\mathrm{disk} + \int_0^\infty \frac{dt}{2\sqrt{\pi t}} \, e^{-t} \, \ln\left|1-2rt\right| \right).
\ee
At hight temperature~$T \gg T_c$,~$r$ is small, so the second term is suppressed like~$\Ocal(r)$ and the free energy is controlled by the disconnected topology.  On the other hand, at low temperature~$T \ll T_c$,~$r$ is large and the integral can formally be expanded in powers of~$1/r$, with the leading contribution given by~$(1/2)\ln r$.  Hence the behavior of the quenched free energy is
\be
\label{eq:CGHSfreeenergyasymptotics}
\overline{F} = -T \begin{cases} 2 \ln (T/T_c) + \cdots, & T \gg T_c \\ \frac{1}{2} \ln (T/T_c) + \cdots, & T \ll T_c \end{cases}.
\ee
where the ellipses denote subleading terms of order unity.  At high temperatures, the free energy is the annealed free energy~$-T \ln \Pcal_\mathrm{disk}$ sensitive only to the the disk topology, while at low temperature the leading-order behavior is modified thanks to the cylinders.

Note that~$\overline{F}$ is still not monotonic in temperature, even with the cylinder contribution.  In particular, while the cylinder contribution decreases the severity of the logarithmic divergence (in reducing the prefactor of~2 to a~$1/2$), it does not eliminate it entirely.  As discussed above, since the calculation of~$\Pcal_m(\beta)$ for integer~$m$ was exact and involved no approximation, the culprit for this unphysical behavior is the analytic continuation away from integer~$m$\footnote{Another option, of course, is that~$\widehat{\mathrm{CGHS}}$ gravity is itself pathological.  But since we are merely using it as a toy model to foreshadow the same sort of behavior that occurs in JT gravity, our main discussion is not enhanced by considering this possibility.}.  This should come as no surprise, as we have already established that the analytic continuation given by~\eqref{eq:PbetamCGHSreal} does not behave correctly for imaginary~$m$; clearly it needs to be modified to remove the pathological behavior entirely.

Nevertheless, the key point is that the replica trick is \textit{required} to see that~$\overline{F}$ receives large corrections from the cylinder topology right around the temperature where the annealed free energy is badly-behaved.  Without properly understanding how the analytic continuation to non-integer~$m$ is to be perfored, we cannot know in precisely what way these additional corrections modify the free energy; the analytic continuation given in~\eqref{eq:PbetamCGHSreal} is insufficient to remove the low-temperature pathology entirely, but we expect that the correct continuation should give a monotonic free energy that yields a vanishing entropy~$-\partial \overline{F}/\partial T$ at zero temperature.  We will revisit this issue in Section~\ref{sec:RSB}.

\section{Free Energy in JT Gravity}
\label{sec:JT}

We have seen that the inclusion of connected topologies in the~$\widehat{\mathrm{CGHS}}$ path integral is of paramount importance for the low-temperature behavior of the free energy.  In that model, the calculation was substantially simplified by the paucity of two-dimensional flat geometries.  We now turn our attention to a more complex gravitational system: JT gravity.

\subsection{Euclidean wormholes can dominate the free energy}

We will first do a preliminary analysis of the role of Euclidean wormholes in the replica computation of the free energy, beginning with a brief review of the salient features of the JT gravity path integral (using specifically the results of Saad, Shenker, and Stanford~\cite{SSS}).  The (Euclidean) JT gravity action is
\be
\label{eq:JTaction}
S_{JT} = -\frac{S_0}{2\pi} \left(\frac{1}{2} \int_M R + \int_{\partial M} K\right) - \left(\frac{1}{2} \int_M \phi(R + 2) + \int_{\partial M} \phi K \right),
\ee
where volume elements are left implied and~$K$ is the extrinsic curvature of~$\partial M$.  When $\partial M$ consists of a single circle, the boundary conditions take the length of~$\partial M$ to be~$\beta/\eps$ and set the dilaton~$\phi|_{\partial M} = \gamma/\eps$ there; after the introduction of an appropriate counterterm, the limit~$\eps \to 0$ is understood.  For simplicity, we will work in units where~$\gamma = 1$; this amounts to working with the dimensionless rescaled inverse temperature and free energy~$\beta/\gamma$,~$\gamma \overline{F}$ respectively.  When~$\partial M$ consists of several circles we may specify boundary conditions separately on each, but for our purposes it will suffice to take all boundary components to have the same length~$\beta/\eps$.

The path integral over the dilaton fixes the path integral over geometries to only include those with constant negative curvature; this space of topologies is of significantly richer structure than its flat counterpart and leads to the organization of the path integral in a genus expansion.  For example, if~$\Pcal_{\mathrm{conn},2}(\beta)$ is the path integral over geometries that connect two boundary components (both of which have length~$\beta/\eps$), pictorially we have
\be
\Pcal_{\mathrm{conn},2}(\beta) = \vcenter{\hbox{\includegraphics[width=0.7\textwidth,page=4]{Figures-pics}}}
\ee
Explicitly, the path integral~$\Pcal_{\mathrm{conn},m}(\beta)$ over geometries that connect~$m$ boundary components is given by
\be
\label{eq:JTgenusexpansion}
\Pcal_{\mathrm{conn},m}(\beta) = \sum_{g = 0}^\infty e^{-S_0(2g+m-2)} Z_{g,m}(\beta),
\ee
where the objects~$Z_{g,m}(\beta)$ are
\begin{subequations}
\label{eqs:Zgm}
\begin{align}
Z_{0,1}(\beta) &= Z_\mathrm{disk}(\beta) \equiv  \frac{e^{2\pi^2/\beta}}{\sqrt{2\pi} \beta^{3/2}}, \\
Z_{0,2}(\beta) &= \int_0^\infty b \, db \, Z_\mathrm{trumpet}(b,\beta)^2 = \frac{1}{4\pi}, \\
Z_{g,m}(\beta) &= \int_0^\infty \left(\prod_{i = 1}^m db_i \, b_i \, Z_\mathrm{trumpet}(b_i,\beta) \right) V_{g,m}(b_1, \ldots, b_m) \mbox{ if } (g,m) \neq (0,1) \mbox{ or } (0,2); \label{subeq:ZgmVgm}
\end{align}
\end{subequations}
here
\be
Z_\mathrm{trumpet}(b,\beta) \equiv \frac{e^{-b^2/(2\beta)}}{\sqrt{2\pi\beta}}
\ee
and~$V_{g,m}(b_1, \ldots, b_m)$ are the volumes of the moduli spaces of Riemann surfaces with~$m$ geodesic boundaries of lengths~$b_1, \ldots, b_m$ (we work in the convention where the normalization~$\alpha$ of these volume forms is one, corresponding to~$V_{0,3} = 1$).  The~$V_{g,m}$ can be computed algorithmically using, for example, Mirzakhani's recursion relation~\cite{Mir06}; a table summarizing the data for small $g$ and $m$ can be found in~\cite{Do11}.

The genus expansion, as well as the contribution of topologies that connect arbitrarily many boundary components, makes the story for JT gravity substantially more involved than for~$\widehat{\mathrm{CGHS}}$.  Nevertheless, even at this heuristic level we can now see that connected topologies must be included in, and will upon inclusion significantly affect,  the low-temperature behavior of the free energy: for example, if we were to only consider the contributions from the disk topology~$Z_{0,1}$ and the ``double trumpet''~$Z_{0,2}$, the analysis would proceed just as in the~$\widehat{\mathrm{CGHS}}$ case, and we would expect the double trumpet contribution to the free energy to compete with that of the disk whenever~$Z_{0,2}/(e^{S_0} Z_{0,1})^2$ is order unity or larger.  For large~$S_0$, this will occur at temperature~$T \lesssim e^{-2S_0/3}$, so that at sufficiently small temperatures failure to include the connected topologies yields a result that is manifestly wrong, as those topologies contribute at least as much as the disconnected ones.

This observation raises a potential concern.  The parameter~$e^{-S_0}$ is supposed to suppress the contributions from higher genus, as well as from topologies that connect more boundary components.  But at low temperature~$\beta \gg 1$, the leading-order behavior of the~$Z_{g,m}$ scales like~$\beta^{(3/2)(2g+m-2)}$, so contributions from higher genus and more-connected topologies are controlled by~$\beta^{3/2} e^{-S_0}$.  The regime in which Euclidean wormholes contribute to the free energy therefore corresponds to the parametric regime in which we lose perturbative control of the genus expansion.  What do we make of this?

From the perspective of the Euclidean wormholes, the story is completely analogous to that of quantum extremal islands in the computation of the entropy of Hawking radiation~\cite{PenShe19,AlmHar19}.  In that case, there is an auxiliary parameter~$k$ parametrizing the entropy of matter fields\footnote{In the end-of-the-world brane model of~\cite{PenShe19},~$k$ is just the number of internal states of the brane.}, and replica wormholes lead to the presence of a quantum extremal island when~$k$ is nonperturbatively large: the Page transition happens at~$k \sim e^{S_0}$.  In the present context, the inverse temperature~$\beta$ plays the role of~$k$.  On the other hand, from the perspective of the genus expansion we are justified in being concerned, because without control of the connected path integral~$\Pcal_{\mathrm{conn},m}$ we cannot expect to make any substantive claim regarding the contribution of Euclidean wormholes.  Fortunately, the regime we are discussing -- that is, taking~$S_0$ large but keeping~$\beta^{3/2} e^{-S_0}$ of order unity -- recovers the so-called Airy case of random matrix integrals, in which the partition function~$Z(\beta)$ is governed by the behavior at the edge of the spectral density~$\rho(E)$.  This simplification makes it possible to resum the genus expansion to include doubly-nonperturbative (in~$S_0$) effects, which we can use to assess how well-behaved the genus expansion is.  Before proceeding, it will therefore be useful to discuss this regime in more detail.

\subsection{The Airy limit}
\label{subsec:Airy}

\begin{figure}[t]
\centering
\subfloat[][]{
\includegraphics[width=0.49\textwidth]{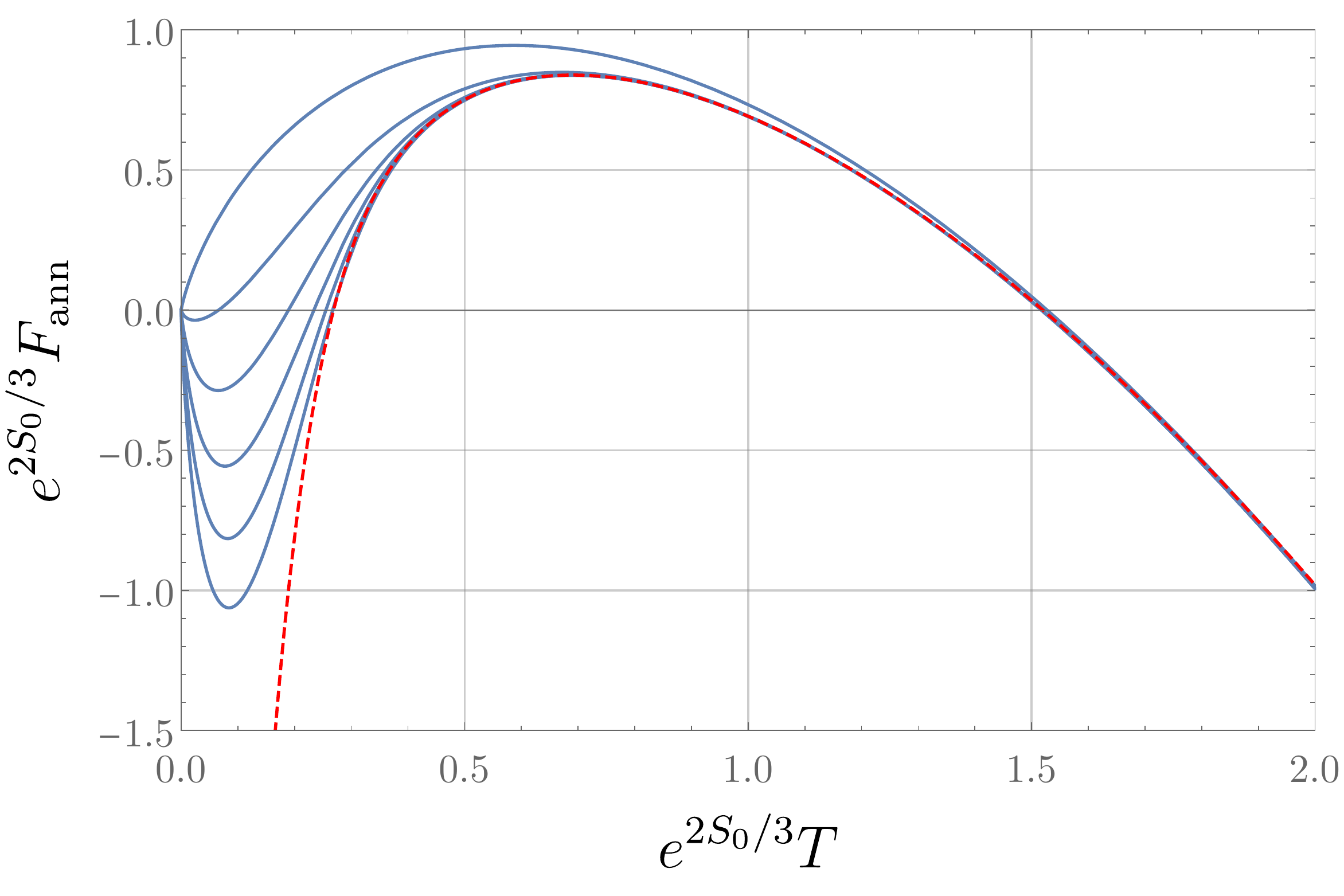}
\label{subfig:JTgenusconverge}
}%
\subfloat[][]{
\includegraphics[width=0.49\textwidth]{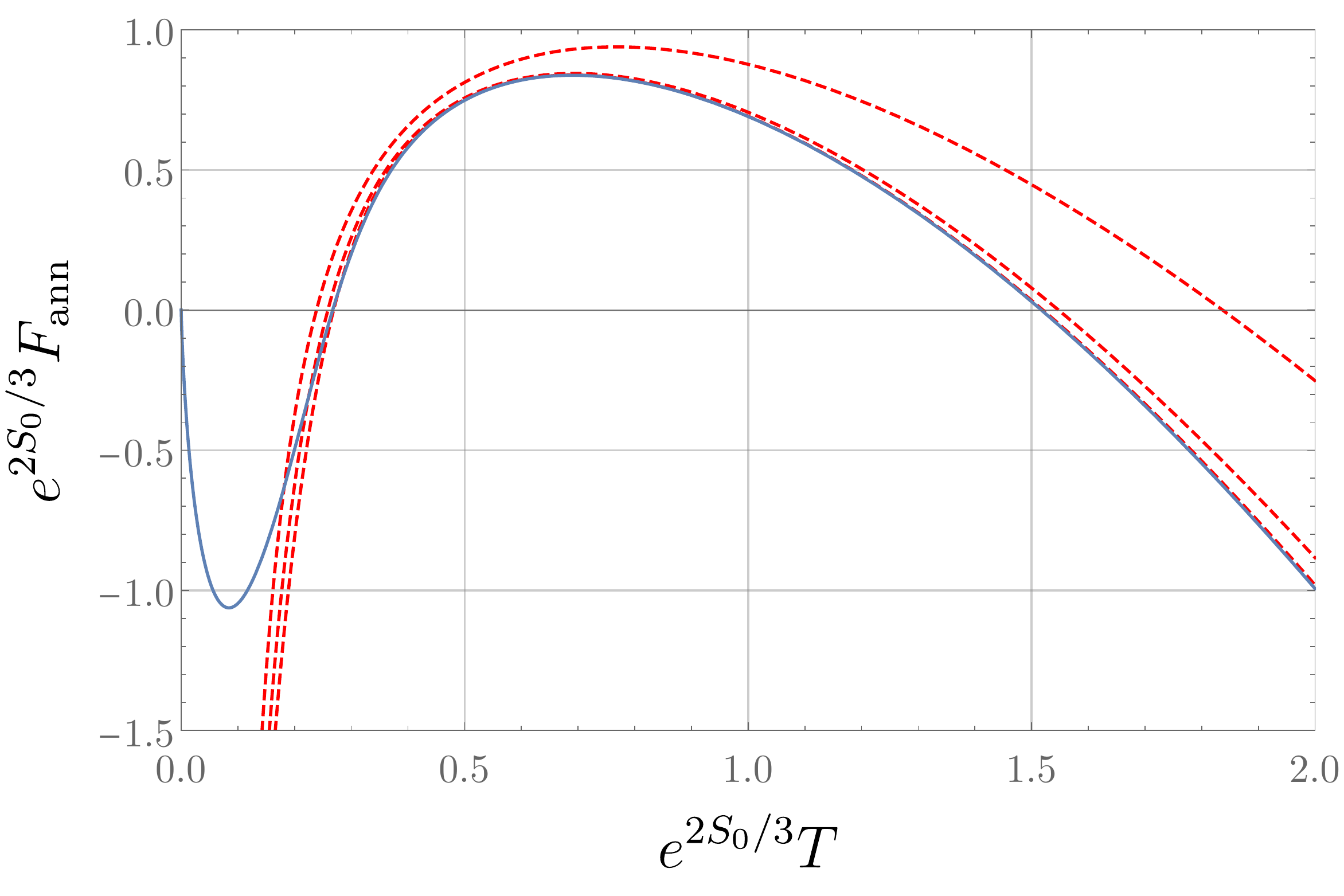}
\label{subfig:JTellconverge}
}
\caption{The annealed free energy~$F_\mathrm{ann}$ for~$S_0 = 7$.  \protect\subref{subfig:JTgenusconverge}: From top to bottom, the solid blue curves show the result after including up to genus~$g = 0,1,2,3,4$, and~5 in the genus expansion~\eqref{eq:JTgenusexpansion}; the dashed red curve shows the result obtained from the low-temperature expansion~\eqref{eq:JTlowtemp} truncated to~$\ell \leq 2$.  \protect\subref{subfig:JTellconverge}: From top to bottom, the dashed red curves show the result after including up to~$\ell = 0,1$, and~2 in the low-temperature expansion~\eqref{eq:JTlowtemp}; the solid blue curve shows the result obtained from keeping up to~$g \leq 5$ in the genus expansion~\eqref{eq:JTgenusexpansion}.  The local maximum at~$e^{2S_0/3} T \approx 0.7$ is robust against the inclusion of higher order perturbative as well as doubly non-perturbative effects.}
\label{fig:JTgenusconverge}
\end{figure}

Before diving into the details of the Airy case\footnote{We are grateful to Douglas Stanford for comments that led to the development of this section.}, let us first do a rough analysis of the behavior of the genus expansion in the regime~$\beta \sim e^{2S_0/3}$ where we expect contributions from Euclidean wormholes to become important.  Recall that the genus expansion~\eqref{eq:JTgenusexpansion} is asymptotic, meaning that it does not converge even when~$\beta^{3/2} e^{-S_0}$ is small.  
Nevertheless, as with any asymptotic series, the partial sums in the genus expansion can be used to bound the free energy.
When $\beta^{3/2} e^{-S_0}$ is not too small, the genus expansion can still be ``under control'' in the sense that the first few
terms in the series~\eqref{eq:JTgenusexpansion} decrease,
so that the partial sums provide a tight bound on the free energy.
To that end, using~\eqref{eqs:Zgm} and the explicit forms of~$V_{g,m}$ found in e.g.~Appendix~B of~\cite{Do11}, in Figure~\ref{subfig:JTgenusconverge} we plot the annealed free energy~$F_\mathrm{ann} \equiv -T \ln \Pcal_{\mathrm{conn},1}$ (corresponding to the disconnected topology free energy~$-T \ln \overline{Z}$) for~$S_0 = 7$ where we include topologies only up to genus~$g = 5$.  The first few partial sums of the genus expansion do indeed provide accurate approximations to the free energy for~$T e^{2S_0/3} \gtrsim 0.3$, which crucially includes a local maximum.  This is suggestive that this maximum should also be present in a full nonperturbative computation of ~$F_\mathrm{ann}$ 
-- but as discussed above, such a maximum is an unphysical feature of the free energy, which we expect to be resolved by the inclusion of connected topologies, 
indicating that inclusion of the latter is indeed necessary.

To proceed more carefully, we can in fact exchange the asymptotic genus expansion for an asymptotic low-temperature expansion with~$T e^{2S_0/3}$ fixed, verifying that it reproduces the behavior exhibited in Figure~\ref{fig:JTgenusconverge}.  To do so, note that the Weil-Petersson volume forms~$V_{g,m}$ appearing in~\eqref{eqs:Zgm} are polynomials in the~$b_i$, and therefore the~$Z_{g,m}$ are polynomials in~$\beta$ of order~$(3/2)(2g+m-2)$, as mentioned above:
\be
Z_{g,m}(\beta) = \left(\beta^{3/2}\right)^{2g+m-2} \sum_{\ell = 0}^{\infty} \beta^{-\ell} P_{\ell,g,m},
\ee
where (up to various constants) the leading-order terms~$P_{0,g,m}$ are the intersection numbers of Chern classes (more generally, the~$P_{\ell,g,m}$ are intersection numbers of the first Miller-Morita-Mumford class with Chern classes~\cite{DijWit18,OkuSak19}; more explicit expressions can be found in Appendix~\ref{app:Airy}).  Inserting this expression into~\eqref{eq:JTgenusexpansion}, for certain~$m$ the sum over genus can be performed as described in~\cite{OkuSak19,OkuSak20} to produce a low-temperature asymptotic expansion; for example, for~$m = 1$ we have
\be
\label{eq:JTlowtemp}
\Pcal_{\mathrm{conn},1}(\beta) = \frac{\exp\left(e^{-2S_0} \beta^3/24\right)}{\sqrt{2\pi} \beta^{3/2}} \, e^{S_0} \sum_{\ell = 0}^\infty \frac{1}{\ell!} \left(\frac{\beta}{2\pi^2}\right)^{-\ell} \tilde{z}_\ell\left(\frac{\beta^{3/2} e^{-S_0} }{\sqrt{2}}\right),
\ee
where the first few~$\tilde{z}_\ell(h)$ are given explictly in~\cite{OkuSak19}.  For~$\beta e^{-2S_0/3}$ of order unity, this asymptotic expansion is under control for large~$\beta$.  In Figure~\ref{subfig:JTellconverge} we show the annealed free energy computed using~\eqref{eq:JTlowtemp} for~$S_0 = 7$, and find that as expected, the low-temperature expansion agrees with the first few partial sums of the genus expansion in the region~$T e^{2S_0/3} \gtrsim 0.3$.   This allows us to conclude that the unphysical peak in the free energy at~$T e^{2S_0/3} \approx 0.7$ cannot be eliminated by either higher order terms in the genus expansion or by doubly non-perturbative effects.

In fact, there is more we can say in this low-temperature limit.  Since~$\tilde{z}_0(h) = 1$, the leading-order term in~\eqref{eq:JTlowtemp} is given by
\be
\Pcal_{\mathrm{conn},1}(\beta) = \frac{\exp\left(e^{-2S_0} \beta^3/24\right)}{\sqrt{2\pi} \beta^{3/2}} \, e^{S_0} + \cdots.
\ee
This is precisely the partition function in the Airy case of random matrix theory and topological gravity,
\be
\overline{Z(\beta)} = \int dE \, \overline{\rho}_\mathrm{Airy}(E) e^{-\beta E} = \frac{\exp\left(e^{-2S_0} \beta^3/24\right)}{\sqrt{2\pi} \beta^{3/2}} \, e^{S_0},
\ee
where the Airy density of eigenvalues is given by~\cite{Wit90,Kon92}
\be
\overline{\rho}_\mathrm{Airy}(E) = e^{2S_0/3} \left[\mathrm{Ai}'\!\left(-e^{2S_0/3} E\right) + e^{2S_0/3} E \, \mathrm{Ai}\left(-e^{2S_0/3} E\right)^2 \right].
\ee
The leading-order behavior (in~$e^{-S_0}$) of~$\overline{\rho}_\mathrm{Airy}(E)$ is just
\be
\label{eq:spectraledge}
\rho_0(E) = \frac{e^{S_0}}{\pi} \, \sqrt{E} \mbox{ with } E > 0,
\ee
which is the universal behavior of the leading-order density of eigenvalues near the edge of the of the spectrum in the double-scaled matrix models of~\cite{SSS}.  Hence the low-temperature expansion~\eqref{eq:JTlowtemp} can be thought of as an expansion about the low-energy edge of the spectrum, with the subleading terms capturing deviations from the exact form~\eqref{eq:spectraledge}.  Concretely, it corresponds to taking~$S_0 \to \infty$ while keeping~$\beta e^{-2S_0/3}$ fixed.  The contribution to~$\Pcal_{\mathrm{conn},m}$ from this leading-order low-temperature behavior can be summed over genus for any~$m$ using the results of~\cite{Oko01}; we summarize the relevant results in Appendix~\ref{app:Airy}, and the relevant expression for~$\Pcal_{\mathrm{conn},m}$ is given by~\eqref{eq:PmAiry}.

The fact that the low-temperature limit in which we are interested is dominated by the universal behavior~\eqref{eq:spectraledge} means that we may gain some qualitative insights into the competition between connected and disconnected topologies by considering particularly simple matrix models.  For example, the Gaussian matrix integral has a leading-order density of eigenvalues given by the Wigner semicircle
\be
\rho_0(E) = \frac{e^{S_0}}{\pi} \sqrt{\frac{a^2 - E^2}{2a}}, \mbox{ with } -a < E < a,
\ee
which recovers~\eqref{eq:spectraledge} in the double-scaling limit~$E \to E - a$ followed by~$a \to \infty$~\cite{GinMoo93}.  The exchange of dominance between connected and disconnected topologies in the Gaussian matrix integral was studied in~\cite{Oku19}, where it was found that the connected correlator~$\overline{Z(\beta)^2}_\mathrm{conn}$ becomes larger than the disconnected correlator~$\overline{Z(\beta)}^2$ at temperatures lower than~$\sim N^{-2/3}$ (or~$\sim e^{-2S_0/3}$ using JT terminology).  So the behavior we are exploring is a general feature of random matrix models.

The upshot is that the low-temperature regime in which we are interested is quite well-understood; importantly, the contributions of higher genera (and their associated doubly-nonperturbative corrections) are insufficient to eliminate the pathological behavior of the annealed free energy.  Therefore, we now turn to a computation of the quenched free energy via an analytic continuation to near~$m = 0$.

\subsection{The continuation in $m$}

To compute the contribution of Euclidean wormholes to the quenched free energy via the replica trick, we need the JT gravitational path integral~$\Pcal_m(\beta)$ defined by~$m$ disconnected boundary circles, each of length~$\beta/\eps$.  These are related to the connected path integrals~\eqref{eq:JTgenusexpansion} by the usual relation
\be
\label{eq:connectedexpansion}
\sum_{m = 0}^\infty \frac{t^m}{m!} \Pcal_m(\beta) = \exp\left(\sum_{m = 1}^\infty \frac{t^m}{m!} \Pcal_{\mathrm{conn},m}(\beta)\right).
\ee
In order to continue to near~$m = 0$, we need to express~$\Pcal_m(\beta)$ in a form analytic in~$m$; this is difficult because the Weil-Petersson volume forms~$V_{g,m}$, and consequently the coefficients~$Z_{g,m}(\beta)$ in the genus expansion, are not known analytically in~$m$.  This is true also in the Airy limit discussed in Section~\ref{subsec:Airy} where although explicit formulas are known (see Appendix A for a review) they are not written as analytic functions of $m$.  We will therefore proceed in an alternative fashion: we define a ``truncated'' path integral~$\Pcal_{m,M}$ to be the JT gravity path integral including only topologies that connect up to~$M$ boundaries, with~$M$ some fixed integer (this amounts to truncating the sum on the right-hand side of~\eqref{eq:connectedexpansion} to~$m \leq M$).  We then analytically continue~$\Pcal_{m,M}$ to non-integer~$m$ with~$M$ held fixed, defining a truncated free energy
\be
\overline{F}_M = - T \lim_{m \to 0} \frac{1}{m} \left(\Pcal_{m,M}(\beta) - 1\right).
\ee
Now, for integer~$m \leq M$,~$\Pcal_{m,M}(\beta)$ will of course coincide with the exact result~$\Pcal_m(\beta)$, and hence for all integer~$m$ we have
\be
\Pcal_m(\beta) = \lim_{M \to \infty} \Pcal_{m,M}(\beta).
\ee
If as~$M \to \infty$ the analytic continuation of~$\Pcal_{m,M}(\beta)$ to non-integer~$m$ converges to a function~$\Pcal_{m,\infty}(\beta)$ which is also analytic in~$m$, we may take~$\Pcal_{m,\infty}(\beta)$ to define the analytic continuation of~$\Pcal_m(\beta)$ to non-integer~$m$.  We can then express the free energy as\footnote{Assuming the limits~$M \to \infty$,~$m \to 0$ commute.}
\be
\overline{F} = \lim_{M \to \infty} \overline{F}_M = -T \lim_{M \to \infty} \lim_{m \to 0} \frac{1}{m} \left(\Pcal_{m,M}(\beta) - 1\right).
\ee
In practice, we will compute the truncated free energies~$\overline{F}_M$ for some relatively small values of~$M$, which by the argument above we might expect to give us an approximation to the exact free energy~$\overline{F}$.  In particular,~$\overline{F}_1$ is just the annealed free energy shown in Figure~\ref{fig:JTgenusconverge}, so we are interested in modifications to the behavior of~$\overline{F}_M$ as~$M$ is increased, specifically in the regime~$T e^{2S_0/3} \gtrsim 0.3$.

To obtain the aforementioned continuation of~$\Pcal_{m,M}(\beta)$ to non-integer~$m$, we proceed inductively: noting that for~$M = 1$ we have~$\Pcal_{m,1}(\beta) = \Pcal_{\mathrm{conn},1}(\beta)^m$, we will suppose that for arbitrary~$M$ we may write
\be
\label{eq:PmMinductive}
\Pcal_{m,M}(\beta) = \, ^{(M)} \! \sum_I \left(A_I^{(M)}\right)^m
\ee
for some~$m$-independent object~$A_I^{(M)}$, where the sum~$^{(M)} \! \sum_I$ (and the corresponding index~$I$) is very schematic and can include both discrete sums and integrals.  We then show that if~$\Pcal_{m,M-1}$ can be written in the form~\eqref{eq:PmMinductive}, then so can~$\Pcal_{m,M}$; since~\eqref{eq:PmMinductive} is true for~$M = 1$, we conclude it holds for all~$M$.  Explicit forms for~$^{(M)} \sum_I$ and~$A_I^{(M)}$ can then be generated by iterating the inductive step.  The continuation of~\eqref{eq:PmMinductive} to non-integer~$m$ is immediate, and the free energy can then easily be obtained.

To perform the inductive step, we wish to express~$\Pcal_{m,M}(\beta)$ as a sum over all possible ways of connecting~$m$ boundaries using topologies that connect no more than~$M$ of them.  To do so, we first choose precisely~$M m'$ of the boundaries to be filled in by wormholes that connect exactly~$M$ boundaries (there will be~$m'$ such wormholes), while the remaining~$m - Mm'$ boundaries will be filled in by topologies connecting no more than~$M-1$ boundaries.  The~$m'$ wormholes connecting the~$Mm'$ boundaries will make a contribution of~$\Pcal_{\mathrm{conn},M}^{m'}$ to the path integral, while the remaining boundaries contribute~$\Pcal_{m-Mm',M-1}(\beta)$.  The full path integral~$\Pcal_{m,M}(\beta)$ is then obtained by summing over all possible~$m'$.  For example, we would pictorially express~$\Pcal_{12,4}$ as
\be
\vcenter{\hbox{\includegraphics[width=0.85\textwidth,page=5]{Figures-pics}}} \, ,
\ee
where dotted lines denote boundaries that contribute to the indicated path integral~$\Pcal_{m,M}$, and each term in the sum should come with a factor that counts how many distinct ways there are of arranging the twelve boundaries into the corresponding configuration.  For general~$m$,~$M$, we have
\be
\Pcal_{m,M}(\beta) = \sum_{m' = 0}^{\lfloor m/M \rfloor} \text{(counting factor)} \Pcal_{\mathrm{conn},M}(\beta)^{m'} \Pcal_{m-Mm',M-1}(\beta),
\ee
where the counting factor is given by
\be
\text{(counting factor)} = \begin{pmatrix} m \\ M m' \end{pmatrix} \times  \frac{1}{m'!} \prod_{j = 1}^{m'} \begin{pmatrix} j M \\ M \end{pmatrix} = \begin{pmatrix} m \\ M m' \end{pmatrix} \frac{(M m')!}{(M!)^{m'} m'!}.
\ee
The first term in this expression simply counts how many distinct ways there are of choosing~$Mm'$ boundaries from the full set of~$m$.  The second term counts how many distinct ways there are of grouping the~$M m'$ boundaries into groups of~$M$; the product over binomial coefficients can be interpreted as the number of ways of choosing~$M$ boundaries to connect out of the total~$m' M$, multipled by the number of ways of choosing~$M$ boundaries out of the remaining~$(m'-1)M$, and so on, with the~$m'!$ cancelling out the overcounting of the same groupings in different orders.  Invoking the inductive hypothesis~\eqref{eq:PmMinductive}, we therefore have
\be
\label{eq:PmMhalfstep}
\Pcal_{m,M}(\beta) = \, ^{(M-1)} \! \sum_I \left(A_I^{(M-1)}\right)^m \sum_{m' = 0}^{\lfloor m/M \rfloor} \begin{pmatrix} m \\ M m' \end{pmatrix} \frac{(M m')!}{m'!} \left(\frac{\Pcal_{\mathrm{conn},M}(\beta)}{M! \left(A_I^{(M-1)}\right)^M}\right)^{m'}.
\ee

We now write
\be
\label{eq:factorialintegrals}
(Mm')! = \int_0^\infty dt \, e^{-t} t^{Mm'}, \qquad \frac{1}{m'!} = \frac{1}{2\pi i} \int_C dz\, e^z z^{-(m'+1)},
\ee
where~$C$ is any contour that encloses~$z = 0$.  Both of these equations are correct for integer~$m'$; for~$m'$ not an integer, the first expression is of course just the definition of the gamma function~$\Gamma(Mm'+1)$ (for~$\mathrm{Re}(Mm') > -1$), but due to the branch cut of~$z^{-(m'+1)}$ along the negative real axis, the second only coincides with~$1/\Gamma(m'+1)$ if~$C$ is chosen to be a Hankel countour\footnote{That is, if~$C$ runs from~$z = -\infty$ to~$z = 0$ and back to~$z = -\infty$, looping in the positive direction around the branch cut.}.  But since~\eqref{eq:factorialintegrals} are only required to hold when~$m'$ is a positive integer, there is no need to require~$C$ to be a Hankel contour, and in the freedom in choosing~$C$ we already see a foreshadowing of the freedom that will manifest in the analytic continuation to near~$m = 0$.

Using the identity~\eqref{eq:identity}, we may evaluate the sum over~$m'$ to obtain
\be
\Pcal_{m,M}(\beta) = \frac{1}{M} \int d\mu(t,z) \sum_{j = 0}^{M-1} \, ^{(M-1)}\sum_I \left(A_I^{(M-1)} + e^{2j\pi i/M} \left(\frac{\Pcal_{\mathrm{conn},M}(\beta)}{M! \, z}\right)^{1/M}  t \right)^m,
\ee
where
\be
d\mu(t,z) \equiv \frac{dt \, dz}{2\pi i z} \, e^{-t+z}
\ee
and the appropriate contours of integration for~$t$ and~$z$ are understood.  This expression for~$\Pcal_{m,M}(\beta)$ is of the form~\eqref{eq:PmMinductive} we assumed for our inductive argument, so we have concluded that~\eqref{eq:PmMinductive} is consistent, with~$A^{(M)}_I$ and the schematic sum~$^{(M)} \! \sum_I$ obeying
\bea
^{(M)} \sum_I &= \int d\mu(t,z) \frac{1}{M} \sum_{j = 0}^{M-1} \, ^{(M-1)} \! \sum_J, \\
A_I^{(M)} &= A_I^{(M-1)} + e^{2j\pi i/M} \left(\frac{\Pcal_{\mathrm{conn},M}(\beta)}{M! \, z}\right)^{1/M}  t.
\eea
Iterating these from the base case~$M = 1$ (for which the sum~$^{(1)} \! \sum_I$ is empty and~$A^{(1)} = \Pcal_{\mathrm{conn,1}}$), we therefore find
\begin{subequations}
\begin{multline}
\label{eq:PgeneralNJT}
\Pcal_{m,M}(\beta) = \int \left(\prod_{k = 1}^{M-1} d\mu(t_k,z_k)\right) \\ \times \frac{1}{M!} \sum_{j_1 = 0}^1 \sum_{j_2 = 0}^2 \cdots \sum_{j_{M-1} = 0}^{M-1} A^{(M)}_{j_1, \ldots, j_{M-1}}(z_1,t_1, \ldots, z_{M-1},t_{M-1})^m,
\end{multline}
\be
\label{eq:ANclosedform}
A^{(M)}_{j_1, \ldots, j_{M-1}}(\{t_k,z_k\}) = \Pcal_{\mathrm{conn},1}(\beta) + \sum_{k = 2}^M e^{2j_{k-1} \pi i/k} \left(\frac{\Pcal_{\mathrm{conn},k}(\beta)}{k! \, z_{k-1}}\right)^{1/k}t_{k-1}.
\ee
\end{subequations}

The analytic continuation to near~$m = 0$ is now straightforward; bearing in mind that as in the~$\widehat{\mathrm{CGHS}}$ case we must take the real part, we find
\be
\label{eq:JTfreeenergy}
\overline{F}_M = -T \, \mathrm{Re} \int \left(\prod_{k = 1}^{M-1} d\mu(t_k,z_k)\right) \frac{1}{M!} \,  \sum_{j_1 = 0}^1 \cdots \sum_{j_{M-1} = 0}^{M-1} \ln  A^{(M)}_{j_1, \ldots, j_{M-1}}(\{t_k,z_k\}).
\ee
As already noted, this free energy depends on the choice of contours~$C_k$ for the integrals over~$z_k$ introduced in the analytic continuation~\eqref{eq:factorialintegrals}.  Specifically, the integrand of~\eqref{eq:JTfreeenergy} exhibits branch cuts in the complex~$z_k$ planes, and will therefore be sensitive to where the contour~$C$ intersects these cuts.  This is not surprising: as discussed in Section~\ref{subsec:CGHScontinuation}, inferring the ``correct'' analytic continuation to near~$m = 0$ is rather subtle.

\begin{figure}[t]
\centering
\subfloat[][Just $g = 0$]{
\includegraphics[width=0.49\textwidth]{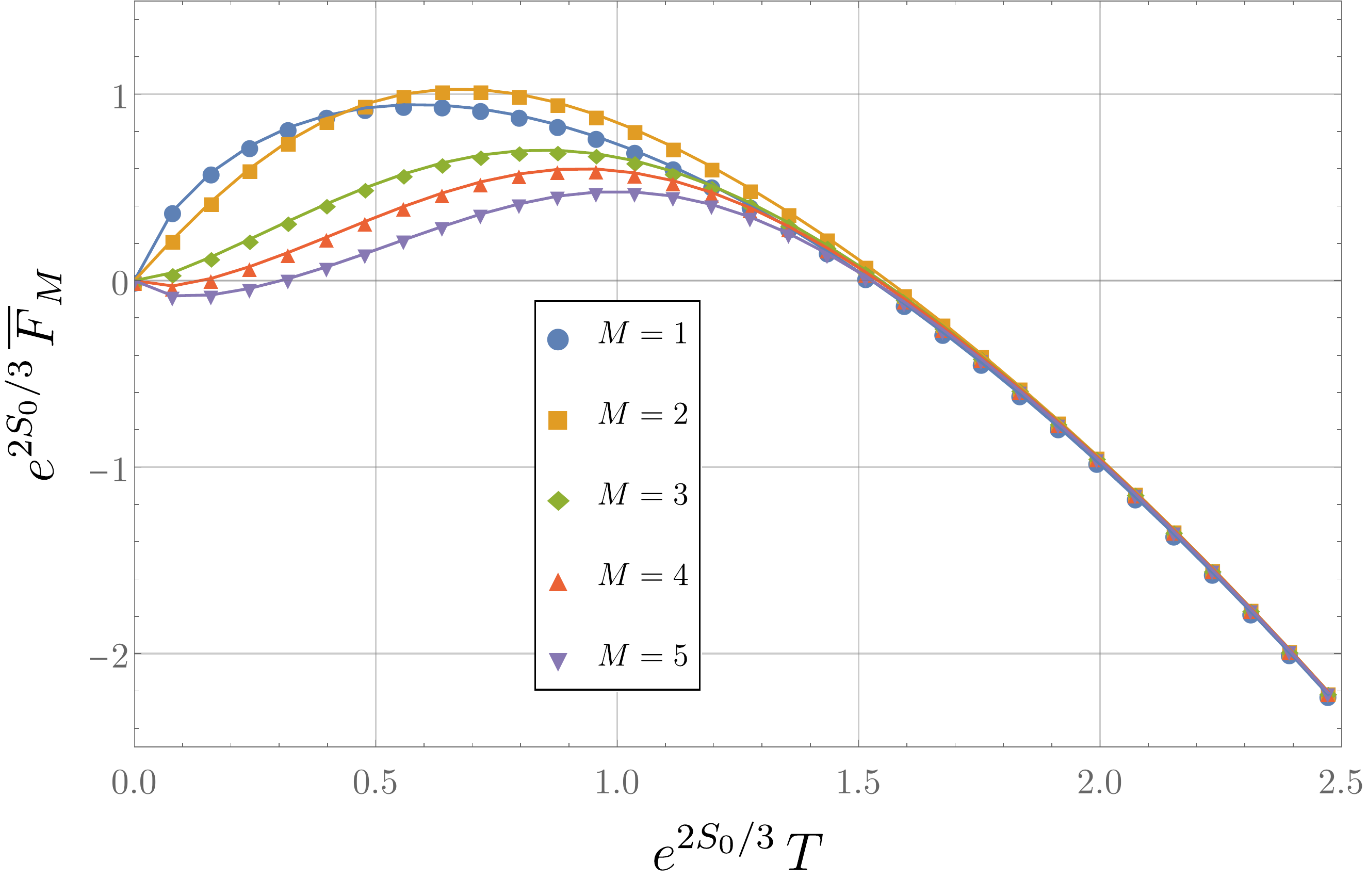}
\label{subfig:JTfreeenergyS0g0second}
}%
\subfloat[][Up to $g = 1$]{
\includegraphics[width=0.49\textwidth]{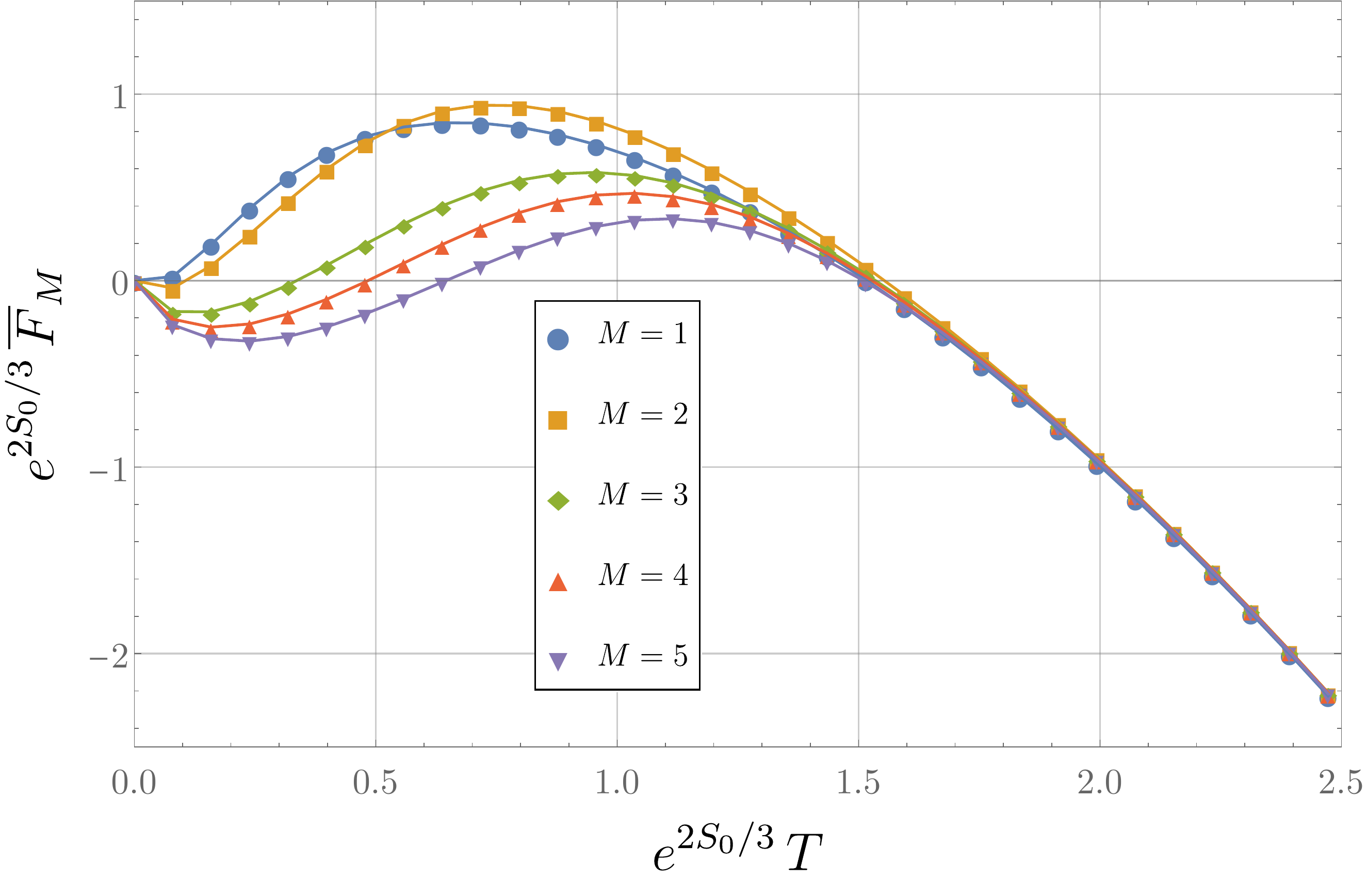}
\label{subfig:JTfreeenergyS0g1second}
}
\\
\subfloat[][Up to $g = 2$]{
\includegraphics[width=0.5\textwidth]{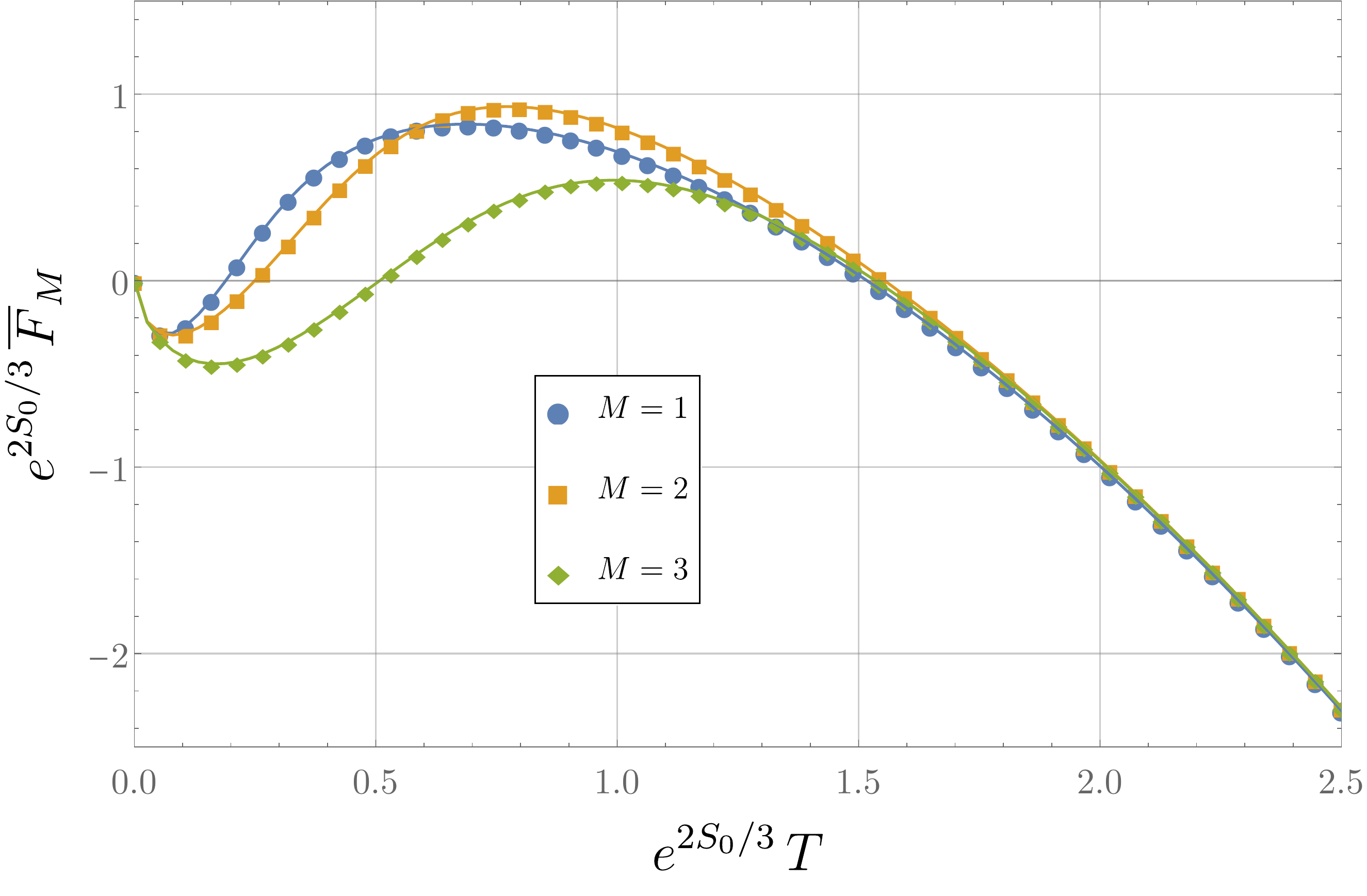}
\label{subfig:JTfreeenergyS0g2second}
}
\caption{The low-temperature behavior of the JT gravity free energy~$\overline{F}_M$ for various~$M$; here we take~$S_0 = 7$, and the contour~$C$ in~\eqref{eq:factorialintegrals} is the unit circle.  From top left to bottom, the energy is computed using topologies with genus up to zero, one, or two.  The blue, orange, green, red, and purple curves correspond to~$M = 1,2,3,4,5$, respectively.}
\label{fig:JTfreeenergyS7}
\end{figure}

\begin{figure}[t]
\centering
\subfloat[][Using~\eqref{eq:factorialintegrals} with~$C$ the unit circle.]{
\includegraphics[width=0.49\textwidth]{JT_Free_Energy_S7_g0_varyM}
\label{subfig:JTfreeenergyS0g0secondcompare}
}%
\subfloat[][Using~\eqref{eq:gammamultiplication}.]{
\includegraphics[width=0.49\textwidth]{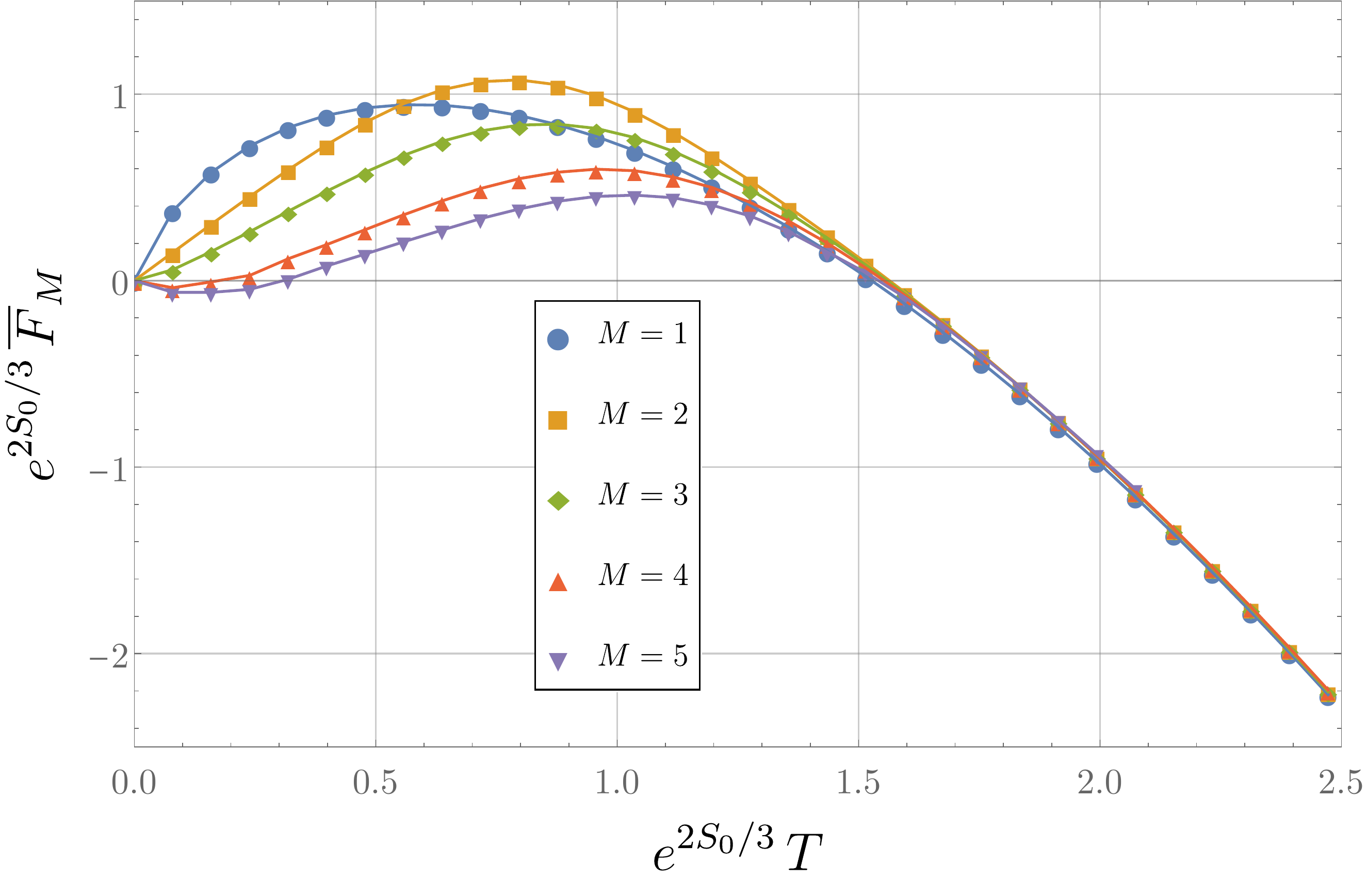}
\label{subfig:JTfreeenergyS0g0firstcompare}
}
\caption{The low-temperature behavior of the JT gravity free energy~$\overline{F}_M$ obtained using two different analytic continuations to non-integer~$m$: on the left we used the continuation~\eqref{eq:factorialintegrals} with the contour~$C$ taken to be the unit circle (this is the same as Figure~\ref{subfig:JTfreeenergyS0g0second}), while on the right we used~\eqref{eq:gammamultiplication}.  The qualitative features agree, but quantitative details do not.  The blue, orange, green, red, and purple curves correspond to~$M = 1,2,3,4,5$, respectively, and we take~$S_0 = 7$.}
\label{fig:othercontinuation}
\end{figure}

We would now like to verify that the corrections from replica wormholes significantly alter and even dominate the behavior of the free energy in the regime~$e^{2S_0/3} T \gtrsim 0.3$ with~$S_0$ large in which we have shown we have perturbative control of the genus expansion.  To that end, we again use~\eqref{eqs:Zgm} (along with the explicit forms of the~$V_{g,m}$) to compute~$\overline{F}_M$, incorporating contributions up to~$g = 2$ and~$M = 5$; the results are shown in Figure~\ref{fig:JTfreeenergyS7}.  Note that in Figure~\ref{fig:JTfreeenergyS7} we take the contour~$C$ in~\eqref{eq:factorialintegrals} to be the unit circle for simplicity.  It is clear that in the regime~$e^{2S_0/3} T \gtrsim 0.3$, the inclusion of replica wormholes can substantially modify the behavior of the free energy.  The unphysical local maximum appears to be ``softened'' by the replica wormholes contribution, though we should be careful not to draw any firm conclusions about the quantitative features of~$\overline{F}_M$ due to the ambiguity in the continuation to near~$m = 0$ (including, for instance, whether the~$M \to \infty$ limit even exists).  In short, we can ascribe meaning to the fact that the free energy changes when replica wormholes are included, but we cannot know its quantitative behavior until we know how to pick the ``right'' continuation.  To highlight this point, in Figure~\ref{fig:othercontinuation} we compare the~$g = 0$ free energies obtained from the analytic continuation~\eqref{eq:factorialintegrals} with~$C$ the unit circle to another analytic continuation in which we instead used the gamma function multiplication theorem to write
\be
\label{eq:gammamultiplication}
\frac{(Mm')!}{m'!} = \frac{M^{Mm'+1/2}}{(2\pi)^{(M-1)/2}} \prod_{k = 1}^{M-1} \Gamma\left(m' + \frac{k}{M}\right),
\ee
and then expressed the gamma functions in the product in their integral form.  The qualitative features of the free energy computed with these two different analytic continuations agree well, but of course they differ quantitatively.  At this point we do not know how to specify the correct prescription, but for reasons that we will describe in the next section, we expect the answer will involve replica symmetry breaking in the~$m \to 0$ limit.

\begin{figure}[t]
\centering
\includegraphics[width=0.49\textwidth]{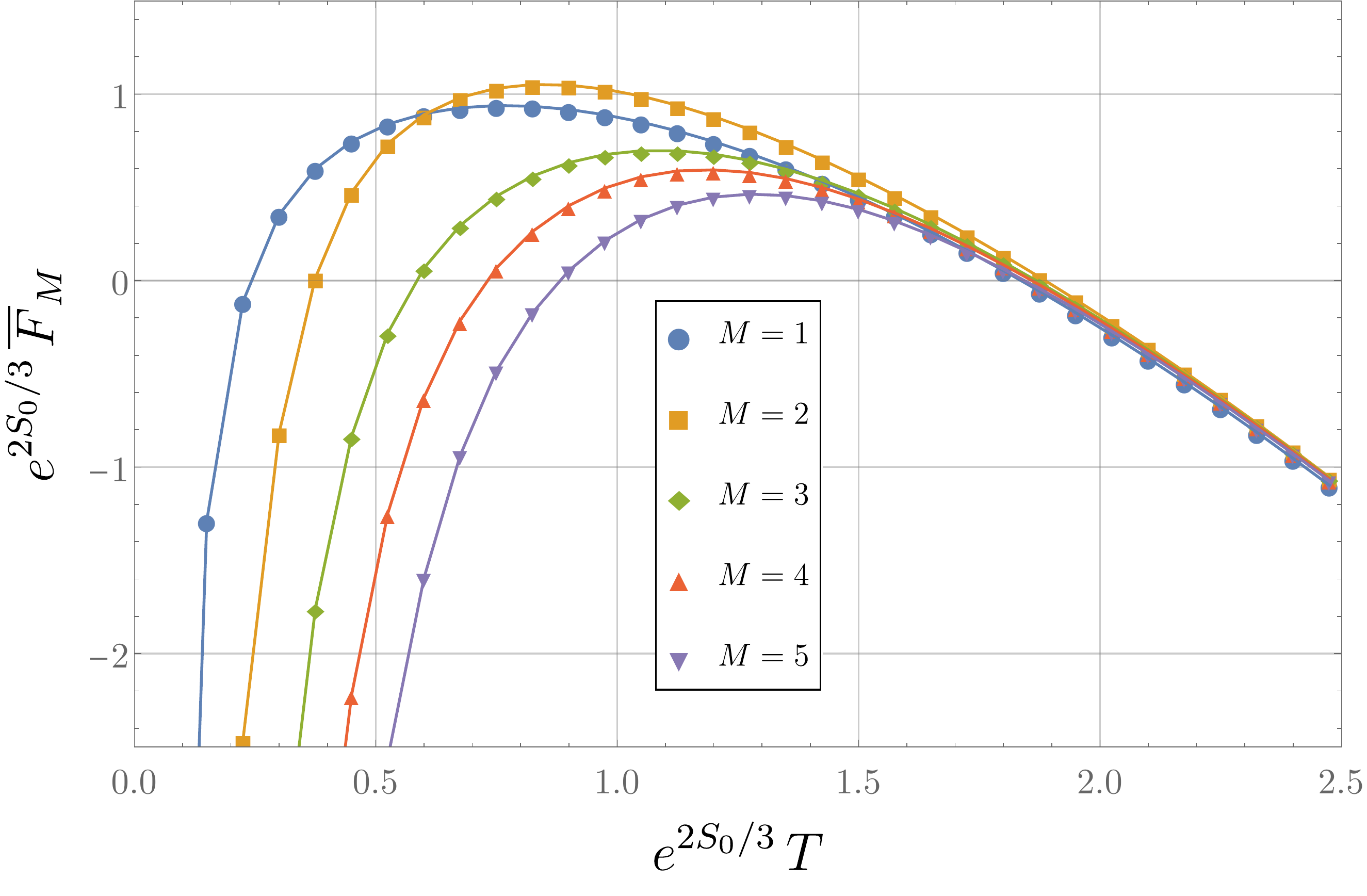}
\caption{The behavior of the quenched free energy~$\overline{F}_M$ for the Airy case, including contributions from all genera using the result~\eqref{eq:PmAiry}.  This amounts to taking~$S_0 \to \infty$ with~$T e^{2S_0/3}$ held fixed in the JT path integral.  As in Figure~\ref{fig:JTfreeenergyS7}, here we take the contour~$C$ in~\eqref{eq:factorialintegrals} to be the unit circle, and the blue, orange, green, red, and purple curves correspond to~$M = 1,2,3,4,5$, respectively.}
\label{fig:Airyfree}
\end{figure}

As a final note, it is interesting to examine the behavior of~$\overline{F}_M$ using the the leading-order low-temperature behavior of~$\Pcal_{\mathrm{conn},m}$ discussed in Section~\ref{subsec:Airy} and Appendix~\ref{app:Airy}.  Specifically, using equation~\eqref{eq:PmAiry} for the path integral in the Airy limit, we obtain the behavior of~$\overline{F}_M$ shown in Figure~\ref{fig:Airyfree}.  While again we may not draw any definitive quantitative conclusions due to the ambiguity in the analytic continuation, we see that connected topologies affect the behavior of the free energy even when all all terms in the genus expansion are included.

\section{Replica Symmetry Breaking and a Spin Glass Analogy}
\label{sec:RSB}

We have shown that in computing extensive quantities like the free energy in gravitational systems, the interpretation of the GPI as an ensemble average -- requiring a replica trick for the computation of the quenched free energy -- can lead to a contribution from replica wormholes that exceeds that of disconnected topologies.  The necessity of these corrections can already be inferred from the pathological properties of the low-temperature behavior of the annealed free energy, computed just from disconnected topologies without resorting to a replica trick.  We have also seen that the inclusion of replica wormholes remedies some of these pathologies but is not sufficient to remove them entirely; we interpret this as necessitating a clearer understanding of the correct analytic continuation to~$m = 0$. Indeed, let us emphasize that in the simpler case of $\widehat{\mathrm{CGHS}}$, the (nonperturbative) calculation that includes all of the allowed geometries \textit{still} exhibits a pathological annealed free energy at low temperatures.  This calculation had only one potential pitfall: the $m\rightarrow 0$ analytic continuation. This immediately implies that it is the  choice of the straightforward analytic continuation that is directly responsible for the incorrect result. 

All of these features -- an annealed free energy with pathological low-temperature behavior, an improvement in this behavior under the inclusion of connected replicas in computing the quenched free energy, and the need for a careful analytic continuation to near~$m = 0$ to eliminate the pathological behavior entirely -- are exhibited in the well-studied context of spin glasses.  In order to draw an analogy with these systems, we will now review one particularly well-known example: the Sherrington-Kirkpatrick (SK) model~\cite{SheKir75}.  In this system, we will see that the non-uniqueness of the analytic continuation to~$m = 0$ is due to a replica symmetry-breaking transition that occurs at~$m < 1$, suggesting that a similar transition likely occurs in the gravitational systems we have examined, and that it is unlike the usual~$\mathbb{Z}_{n}$-replica symmetry breaking that is discussed in the context of gravitational calculations of the Renyi entropies.  We will keep the review of the SK model limited to the bare essentials, but would recommend~\cite{SherringtonReview,CasCavReview} and especially~\cite{SpinGlassBook} for more comprehensive treatments.

\subsection{Review of the SK model}

The SK model is an infinite-ranged classical Ising model of~$N$ interacting spins~$\sigma_i$, with Hamiltonian
\be
H_{\{J_{ij}\}}[\sigma] = -\sum\limits_{(ij)} J_{ij}\sigma_{i}\sigma_{j},
\ee
where the sum runs over all distinct pairs of spins~$(ij)$.  Each of the random couplings~$J_{ij}$ is drawn from a Gaussian\footnote{We could consider a more general distribution, but the important physics is captured by just the second moment of~$P(J_{ij})$.} distribution~$P(J_{ij})$ with mean~$J_0/N$ and and variance~$J^2/N$.  As above, we will denote averages over the distribution~$P(J_{ij})$ via an overline, so that, for instance, the ensemble average of the logarithm of the partition function is
\be
\overline{\ln Z} = \int \left(\prod_{(ij)} dJ_{ij} P(J_{ij}) \right) 
\ln \Tr e^{-\beta H_{\{J_{ij}\}}[\sigma]}.
\ee
Note that~$\overline{\ln Z}$ is quite difficult to compute directly, but using the replica trick~\eqref{eq:replicatrick} requires us to simply compute the ensemble average of the $m$-replicated partition function
\be
\overline{Z^m} = \overline{
\left(\Tr e^{-\beta H_{\{J_{ij}\}}[\sigma]}\right)^m} = \overline{\Tr_m \exp\left(-\beta \sum_{\alpha = 1}^m H_{\{J_{ij}\}}[\sigma^\alpha]\right)},
\ee
where~$\alpha$ is a replica index that labels~$m$ copies of the spins~$\sigma^\alpha$, and the last trace is over all~$m$ replica systems.  The last average is quite easy to express in terms of the moments~$J_0$ and~$J$ of the distribution~$P(J_{ij})$:
\be
\label{eq:ZmSKdisorderaverage}
\overline{Z^m} = \Tr_m \exp \left\{\frac{1}{N} \sum_{(ij)} \left(J_0 \beta \sum_{\alpha = 1}^m \sigma_i^\alpha \sigma_j^\alpha + \frac{(\beta J)^2}{2} \left(\sum_{\alpha = 1}^m \sigma_i^\alpha \sigma_j^\alpha\right)^2 \right) \right\}.
\ee
The fact that the couplings~$J_{ij}$ are correlated between the different replicas has led to the introduction of an effective coupling between replicas via the ensemble average.  Moreover, by completing the squares in the sums over spin sites and introducing auxiliary variables~$s_\alpha$,~$q_{(\alpha,\gamma)}$ with~$\alpha \neq \gamma$ (sometimes called Hubbard-Stratonovich variables, collective fields, or mean fields), we may decouple the spin sites:
\be
\label{eq:ZmSKexact}
\overline{Z^m} = B \int \left(\prod_\alpha ds_\alpha\right) \left(\prod_{(\alpha,\gamma)} dq_{(\alpha,\gamma)}\right) e^{N H_\mathrm{eff}},
\ee
where~$B$ is a prefactor that is sub-exponential in~$N$ (and therefore will be irrelevant in the thermodynamic limit~$N \to \infty$), the variables~$s_\alpha$ and~$q_{(\alpha,\gamma)}$ are all integrated over the real axis, and the notation~$(\alpha,\gamma)$ denotes all distinct pairs of replicas.  Here the effective Hamiltonian~$H_\mathrm{eff}$ is independent of~$N$ and given by
\begin{subequations}
\be
H_\mathrm{eff} = \ln \underset{\{\sigma^\alpha\}}{\Tr} e^{\Lcal[\sigma^\alpha]} - \Kcal,
\ee
where the trace is now over all~$m$ replicas of a \textit{single spin site} and
\begin{align}
\Kcal &\equiv \frac{\beta J_0}{2} \sum_\alpha s_\alpha^2 + \frac{(\beta J)^2}{2} \sum_{(\alpha,\gamma)} q_{(\alpha,\gamma)}^2 - \frac{m}{4} (\beta J)^2, \\
\Lcal[\sigma^\alpha] &\equiv \beta J_0 \sum_{\alpha} s_{\alpha} \sigma^{\alpha} + (\beta J)^2 \sum_{(\alpha,\gamma)} q_{(\alpha,\gamma)} \sigma^{\alpha} \sigma^{\gamma}.
\end{align}
\end{subequations}

At this point~\eqref{eq:ZmSKexact} is still an exact equation, whose existence is made possible thanks to the all-to-all coupling of the SK model:  the fact that the couplings between all pairs of sites are drawn from the same distribution allows for the factorization of different spin sites in~\eqref{eq:ZmSKdisorderaverage} via the introduction of the variables~$s_\alpha$ and~$q_{(\alpha,\beta)}$.  We may now take the thermodynamic limit~$N \to \infty$, finding via a saddle point approximation that
\be
\label{eq:ZmSKsaddle}
\overline{Z^m} \sim \exp\left(N H_\mathrm{eff}\left(s_\alpha,q_{(\alpha,\gamma)}\right)\right),
\ee
where now~$s_\alpha$ and~$q_{(\alpha,\gamma)}$ are solutions to the saddle point equations~$\partial H_\mathrm{eff}/\partial s_\alpha = 0 = \partial H_\mathrm{eff}/\partial q_{(\alpha,\gamma)}$.  It is easy to see that these conditions reduce to
\be
\label{eq:meanfields}
s_\alpha = \ev{\sigma^\alpha}_\Lcal, \qquad q_{(\alpha,\gamma)} = \ev{\sigma^\alpha \sigma^\gamma}_\Lcal, \quad \mbox{where} \quad \ev{X}_\Lcal \equiv \frac{\Tr_{\{\sigma^\alpha\}}(X e^{\Lcal[\sigma^\alpha]})}{\Tr_{\{\sigma^\alpha\}}e^{\Lcal[\sigma^\alpha]}},
\ee
giving~$s_\alpha$ and~$q_{(\alpha,\gamma)}$ the interpretation of mean fields fixed by the self-consistency conditions~\eqref{eq:meanfields}.  Importantly, the field~$q_{(\alpha,\gamma)}$ is interpreted as a coupling between replicas; a saddle with nonzero~$q_{(\alpha,\gamma)}$ indicates the spontaneous ``turning on'' of this coupling.  Because this coupling is our main focus, from now on we will set~$J_0 = 0$ so that~$\overline{Z^m}$ becomes independent of the mean field~$s_\alpha$ (this excludes the possibility of a ferromagnetic phase, in which we are not currently interested).

In order to now compute~$\overline{\ln Z}$ (and therefore~$\overline{F}$) in the thermodynamic limit, we must analytically continue~\eqref{eq:ZmSKsaddle} to non-integer~$m$ near zero.  Because the sums in~$H_\mathrm{eff}$ are only well-defined for integer~$m$, this procedure requires positing some ansatz for the matrix~$q_{(\alpha,\gamma)}$ that is amenable to the analytic continuation to~$m = 0$.  Given the replica symmetry of the problem (corresponding to the permutation group~$\mathbb{S}_m$), it is natural to take the replica-symmetric ansatz
\be
\label{eq:replicasymmetricq}
q_{(\alpha,\gamma)} = q.
\ee
Indeed, for positive integer~$m$, the dominant saddles do exhibit this symmetry~\cite{HemmenPalmer79}.  The analytic continuation to near~$m = 0$ is then straightforward, and the free energy becomes
\bea
\label{subeq:SKfreeenergy}
-\beta N^{-1} \overline{F} &= \frac{(\beta J)^2}{4} (1-q)^2 + \int_{-\infty}^\infty \frac{dy}{\sqrt{2\pi}} \, e^{-y^2/2} \ln \left(2 \cosh(\beta J \sqrt{q} \, y)\right), \\
& \mbox{where } q = \int_{-\infty}^\infty \frac{dy}{\sqrt{2\pi}} \, e^{-y^2/2} \tanh^2(\beta J \sqrt{q} \, y).
\eea
When~$\beta J < 1$ (i.e.~at sufficiently high temperature), the only solution is~$q = 0$, and hence the replicas are uncorrelated; this is the paramagnetic phase. The free energy obtained in this phase therefore satisfies $\overline{\ln Z} = \ln \overline{Z}$, i.e. we may average $Z$ before taking the logarithm with no loss of information. Hence the replica trick does not introduce any novel behavior.  As the temperature is lowered, however, a solution with nonzero~$q$ begins to exist once~$\beta J > 1$.  This new solution dominates the free energy\footnote{The number of off-diagonal components of~$q_{(\alpha,\gamma)}$ is~$m(m-1)/2$, which is \textit{negative} for~$0 < m < 1$; this implies that the saddle that maximizes~$\overline{Z^m}$ with with respect to the components~$q_{(\alpha,\gamma)}$ actually \textit{minimizes}~$\overline{Z^m}$ with respect to~$q$ when~$m < 1$.  Hence the saddle that dominates the free energy is in fact the one that \textit{maximizes} it with respect to~$q$. \label{foot:maximize}}, corresponding to the spin-glass phase in which the replicas spontaneously couple.

While the field~$q$ was introduced in the context of the replica formalism, it has an interpretation in the~$m \to 0$ limit: it computes the so-called Edwards-Anderson order parameter~$q_\mathrm{EA}$ defined by the disorder-averaged square magnetization~\cite{EdwAnd75}:
\be
\lim_{m \to 0} q = q_\mathrm{EA} \equiv \overline{\ev{\sigma_i}^2}.
\ee
Here independence of the choice of lattice site~$i$ follows from translational invariance (after the disorder average), and the expectation value is a standard thermodynamic average taken with respect to a particular sampling of couplings:
\be
\ev{\sigma_i} \equiv \frac{\Tr \sigma_i e^{-\beta H_{\{J_{ij}\}}}}{\Tr e^{-\beta H_{\{J_{ij}\}}}}.
\ee
The non-vanishing of~$q$ in the spin glass phase therefore corresponds to magnetic order for any particular sampling of the couplings~$J_{ij}$.  However, for~$J_0 = 0$ the disorder-averaged magnetization vanishes:~$\overline{\ev{\sigma_i}} = 0$.  Since this disorder-averaged magnetization measures the ferromagnetic order of the system, we see that the spin-glass phase corresponds to a cooperatively frozen magnetic state but with no ferromagnetic order.

\subsection{Replica symmetry breaking in the SK model}

\begin{figure}[t]
\centering
\includegraphics[width=0.5\textwidth]{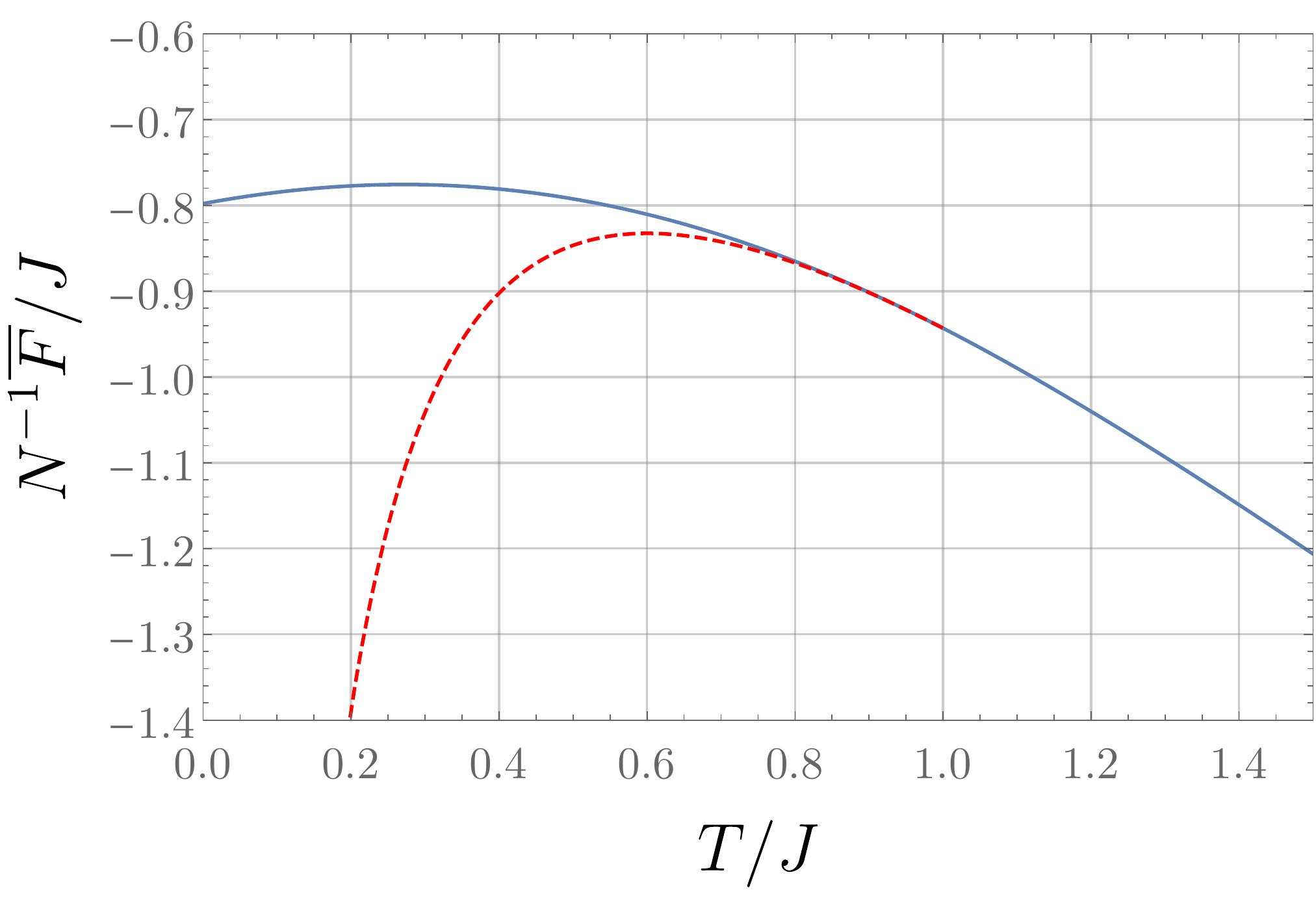}
\caption{The free energy of the SK model, computed using the replica-symmetric ansatz~\eqref{subeq:SKfreeenergy}.  For~$T/J > 1$, there is only the paramagnetic phase~$q = 0$; continuing this phase to~$T = 0$ (dashed red line) gives a free energy that is non-monotonic and divergent at~$T = 0$.  The appearance of the spin glass phase~$q \neq 0$ when~$T/J < 1$ (solid blue line) removes the divergence, but the free energy is still non-monotonic.  (As mentioned in footnote~\ref{foot:maximize}, a feature of the analytic continuation to~$m = 0$ is that the dominant phase is in fact the one that \textit{maximizes} the free energy.)}
\label{fig:SKfreeenergy}
\end{figure}

As can be seen directly from~\eqref{subeq:SKfreeenergy}, the free energy of the paramagnetic phase~$q = 0$ is pathological if we  extend it to arbitrarily low temperature: at large temperatures it scales like~$-T$, while at low temperatures is exhibits a~$-1/T$ divergence.  These behaviors imply that it is non-monotonic, with the thermodynamic entropy becoming negative at sufficiently low temperatures (and in fact diverging at zero temperature).  As shown in Figure~\ref{fig:SKfreeenergy}, the turning on of the spin glass phase when~$T/J < 1$ is necessary to alleviate these pathologies, rendering the free energy finite.  However, it is still non-monotonic: the zero-temperature entropy is~$\overline{S}_{T = 0} = -N/2\pi$.  Clearly the calculation remains incomplete; from our earlier discussion, we expect that this missing ingredient involves some nontrivial behavior of the analytic continuation from~$\overline{Z^m}$ at positive integer~$m$ to~$m = 0$\footnote{Though we note that unlike the~$\widehat{\mathrm{CGHS}}$ case discussed in Section~\ref{subsec:CGHScontinuation}, the analytic continuation of the replica-symmetric ansatz~\eqref{eq:replicasymmetricq} in~\eqref{eq:ZmSKsaddle} to imaginary~$m$ does indeed obey the boundedness condition~$\left|\overline{Z^{i\alpha}}\right| \leq 1$.  However,~$\overline{Z^m}$ still exhibits superexponential growth for real~$m$, so Carlson's theorem is still inapplicable~\cite{HemmenPalmer79}.}.  How do we understand what the correct analytic continuation is?

The answer can be gleaned by performing a stability analysis of the replica-symmetric ansatz~\eqref{eq:replicasymmetricq}.  Indeed, though~\eqref{eq:replicasymmetricq} does give the correct form of the saddles for computing~$\overline{Z^m}$ when~$m$ is a positive integer, it becomes unstable for sufficiently small~$m < 1$: an eigenvalue of the Hessian~$\partial^2 H_\mathrm{eff}/\partial q_{(\alpha,\gamma)} \partial q_{(\beta,\delta)}$ evaluated on the ansatz~$q_{(\alpha,\gamma)} = q$ becomes positive in the limit~$m \to 0$~\cite{AlmTho78}.  We must therefore invoke an alternative ansatz for~$q_{(\alpha,\gamma)}$ that avoids this instability as~$m \to 0$.  The correct analytic continuation to~$m = 0$ will then be determined by the behavior of the ansatz for~$q_{(\alpha,\gamma)}$ which remains stable down to~$m = 0$; this behavior will undergo a phase transition at some critical~$m_c(T) < 1$~\cite{Kon83} that was missed by just considering the replica-symmetric ansatz~\eqref{eq:replicasymmetricq}.  The presence of this phase transition means that it is crucial to analytically continue the saddle-point equations~$\partial H_\mathrm{eff}/\partial q_{(\alpha,\beta)} = 0$ themselves down to~$m = 0$, rather than first evaluating their on-shell value at integer~$m$ and then analytically continuing the results.

Because the number of components of~$q_{(\alpha,\gamma)}$ is~$m(m-1)/2 < 0$ when~$m < 1$, it is far from obvious how to construct a replica symmetry-breaking (RSB) ansatz that is amenable to analytic continuation.  The answer is the well-established Parisi ansatz~\cite{Par79a,Par79b,Par80a,Par80b}.  To get an idea of how this procedure works, consider splitting up the~$m$ replicas that define~$\overline{Z^m}$ into groups of~$m_1$, with~$m_1$ an integer that divides~$m$.  We then write~$q_{(\alpha,\gamma)}$ in a block-diagonal form according to this grouping:
\be
\label{eq:1RSB}
q_{(\alpha,\gamma)} = \begin{pmatrix} Q_2 & Q_1 & Q_1 & Q_1 \\ Q_1 & Q_2 & Q_1 & Q_1 \\ Q_1 & Q_1 & Q_2 & Q_1 \\ Q_1 & Q_1 & Q_1 & Q_2 \end{pmatrix},
\ee
where~$Q_1$ and~$Q_2$ are~$m_1 \times m_1$ matrices all of whose entries are~$q_1$ and~$q_2$, respectively (in this example, we have~$m/m_1 = 4$).  This ansatz for~$q_{(\alpha,\beta)}$ can be analytically continued to~$m = 0$ while leaving~$m_1$,~$q_1$, and~$q_2$ free as variational parameters to be fixed by extremizing the free energy with respect to them (since~$1 \leq m_1 \leq m$, the analytic continuation of~$m$ also continues~$m_1$ to be between zero and one).  This procedure, called one-step RSB (or 1RSB), substantially improves the pathologies in the free energy shown in Figure~\ref{fig:SKfreeenergy}, but the zero-temperature entropy is still negative (though substantially closer to zero)\footnote{There are other models of spin glasses in which~1RSB is in fact sufficient to obtain a stable ansatz, e.g.~the~$p$-spin spherical model~\cite{Der80,Der81,CriSom92}.}.

To proceed further, we iterate this procedure: we introduce a new integer~$m_2$ that divides~$m_1$ and partition~$Q_2$ into the same block-diagonal structure as~\eqref{eq:1RSB},
\be
Q_2 = \begin{pmatrix} Q_3 & \widetilde{Q}_2 & \widetilde{Q}_2  \\ \widetilde{Q}_2 & Q_3 & \widetilde{Q}_2 \\ \widetilde{Q}_2 & \widetilde{Q}_2 & Q_3 \end{pmatrix},
\ee
where~$\widetilde{Q}_2$ and~$Q_3$ are~$m_2 \times m_2$ matrices all of whose entries are~$q_2$ and~$q_3$, respectively (in this example~$m_1/m_2 = 3$).  Repeating this process~$r$ times, we may then continue to~$m = 0$, obtaining an expression for the free energy that depends on~$2p+1$ variational parameters,~$m_i$ for~$i = 1, \ldots, p$ and~$q_i$ for~$i = 1, \ldots, p+1$.  After the continuation to~$m = 0$ has been made, we may in fact take the limit~$p \to \infty$ which turns the~$(q_i,m_i)$ into a continuous function~$q(x)$.  The free energy is then a functional of~$q(x)$, and is obtained by a functional extremizaton with respect to~$q(x)$.

One way of understanding what the~$p \to \infty$ limit means is as follows.  For positive integer~$m$, the the ansatz~\eqref{eq:1RSB} breaks the full replica symmetry group~$\mathbb{S}_m$ into the subgroup
\be
\mathbb{S}_m \xrightarrow[\mathrm{break}]{} \left(\mathbb{S}_{m_1}\right)^{\otimes m/m_1} \otimes \mathbb{S}_{m/m_1},
\ee
with the first factor corresponding to the permutation symmetry of each of the groups of~$m_1$ rows and columns, and the second corresponding to the permutation symmetry of the~$m/m_1$ groups amongst themselves.  The iterative procedure outlined above amounts to breaking the subgroup further, into
\be
\mathbb{S}_m \xrightarrow[\mathrm{break}]{} \mathbb{S}_{m/m_1} \otimes \bigotimes_{i = 1}^{p} (\mathbb{S}_{m_i/m_{i+1}})^{\otimes m/m_i}
\ee
(with~$m_{p+1} \equiv 1$), but of course we cannot take~$p$ arbitrarily large if the~$m_i$ must all be divisors of~$m$.  However, if we analytically continue this group structure to~$m = 0$, we obtain
\be
\mathbb{S}_0 \xrightarrow[\mathrm{break}]{} \mathbb{S}_0 \otimes \bigotimes_{i = 1}^{p} (\mathbb{S}_{m_i/m_{i+1}})^{\otimes 0}.
\ee
So we find that~$\mathbb{S}_0$ contains itself as a subgroup, which means we may continue to break the symmetry as much as desired by breaking the~$\mathbb{S}_0$ factor on the right-hand side.  This is the feature that allows us to take~$p \to \infty$ in the Parisi ansatz after the continuation to~$m = 0$ has been performed.

The point is that RSB is contained in the structure of the Parisi function~$q(x)$: in the replica-symmetric ansatz~\eqref{eq:replicasymmetricq}~$q(x)$ is just a constant~$q$, so nontrivial structure in~$q(x)$ is indicative of RSB.  Because the Parisi ansatz changes the na\"ive analytic continuation to~$m = 0$, we see that RSB is the mechanism reponsible for the phase transition at~$m < m_c(T)$, and it answers the question posed above: how do we correctly continue to~$m = 0$?

\subsection{RSB in Gravity \`a la Spin Glass}
\label{subsec:RSBgravity}

In Sections~\ref{sec:CGHS} and~\ref{sec:JT} we saw that in simple gravitational models, the introduction of replica wormholes alleviated some of the low-temperature pathologies of the disconnected free energy, but it did not remove them entirely; we interpreted this result as the statement that our anaytic continuation to~$m = 0$ (which in the JT gravity case exhibited considerable freedom) was not correct.  Having now reviewed spin glasses, there is quite an obvious analogy: since the paramagnetic and spin glass phases are characterized by correlated and uncorrelated replicas, respectively, we would like to interpret the ``turning on'' of replica wormholes in the gravitational free energy as the onset of spin glass-like behavior.  It is important to note that the analogy will not be literal: perhaps the most important distinction is that a spin glass is a bona fide sharp phase transition that can be seen in the thermodynamic~$N \to \infty$, whereas we did not work in any saddle point approximation in our gravitational models (and in fact, the fact that the temperature at which connected topologies contributed was nonperturbatively small in~$S_0$ suggests that the transition should be invisible to a semiclassical~$S_0 \to \infty$ analysis).  The most relevant paralle we would like to highlight has to do with the all-important analytic continuation: in the spin glass model, a replica symmetric ansatz remedies some low-temperature pathologies of the free energy, but it gives the incorrect analytic continuation, and RSB must be invoked due to a phase transition at small~$m$.  What does this analogy suggest for how to obtain the correct analytic continuation to~$m = 0$ in the gravitational case?

One of the key lessons to draw from the spin glass example is that a na\"ive analytic continuation from the \textit{values} of~$\overline{Z^m}$ for positive integer~$m$ to near~$m = 0$ gives a wrong answer: we must first analytically continue the \textit{saddle point equations} to near~$m = 0$ with an appropriate ansatz, and only then do we solve them for the small-$m$ behavior of~$\overline{Z^m}$.  In Sections~\ref{sec:CGHS} and~\ref{sec:JT}, this is not what we did: we instead expressed the gravitational path integrals~$\Pcal_m(\beta)$ for integer~$m$, and then looked for an analytic continuation to~$m = 0$.  For the same reason as the spin glass, we might expect that in a gravitational theory we must look for RSB saddle points in order to perform the analytic continuation correctly.

Let us first be clear on what we mean by ``replica symmetry breaking''. There is a sense in which we could say that any replica wormhole breaks replica symmetry, since the symmetry group of~$m$ disconnected boundaries is~$\mathbb{S}_m$, which is broken by any gravitational saddle that connects two or more of these boundaries.  But the sort of RSB that appears in the spin glass example, and which we expect to determine the correct analytic continuation to near~$m = 0$, is something more subtle: it is the breaking at~$m < 1$ of a symmetry that is exhibited by the dominant saddles when~$m$ is a positive integer.  For example, if the~$m$-boundary gravitational path integral is dominated by disconnected saddles whenever~$m$ is a positive integer, the symmetry group is indeed~$\mathbb{S}_m$, and we would say that RSB occurs if this group is broken for~$m < 1$.  But if the path integral for positive integer~$m$ is dominated by, say, a connected wormhole with~$\mathbb{Z}_m$ symmetry, we would not say that RSB occurs as~$m \to 0$ unless the~$\mathbb{Z}_m$ is broken for some~$m < 1$.

Now, since in Sections~\ref{sec:CGHS} and~\ref{sec:JT} we did not work in a saddle point approximation, no equations of motion were involved in our calculation.  Hence it is not immediately clear what the analogue of the Parisi procedure might be in this models.  It may instead be easier to consider working in the semiclassical limit of some more general gravitational theory, in which case probing the role of RSB, and computing the correct analytic continuation to near~$m = 0$, requires us to look for a RSB ansatz for a gravitational solution that allows for the continuation of the gravitational equations of motion to near~$m = 0$.  This is still a difficult task, which is a natural starting point for future work.  Instead, let us compare the approach we have in mind in this context with that of the Lewkowykz-Maldacena replica trick used to compute holographic von Neumann entropies~\cite{LewMal13}.  In the latter case, we are required to compute the gravitational path integeral defined by an~$n$-sheeted connected boundary manifold~$B_n$ with~$\mathbb{Z}_n$ symmetry.  Assuming the dominant bulk saddle also exhibits this symmetry, we may quotient the bulk geometry by~$\mathbb{Z}_n$, after which the analytically-continued bulk equations of motion are just those on a manifold with boundary~$B_1$ consisting of a single sheet, except with a conical defect proportional to~$(n-1)$ at the fixed point of the~$\mathbb{Z}_n$ isometry.  For~$n$ near one, the bulk equations of motion can be expanded perturbatively around the smooth geometry with boundary~$B_1$ and no conical defect, and the condition that the equations of motion hold near the (perturbative) conical defect reproduces the Ryu-Takayagani formula for holographic entanglement entropy~\cite{RyuTak06}.  In this context, the ``usual'' notion of RSB is the breaking of the~$\mathbb{Z}_n$ for~$n \neq 1$ -- but of course there is no breaking of replica symmetry for~$n = 1$, since~$\mathbb{Z}_1$ is trivial.  According to the alternative definition of RSB that occurs in spin glasses, RSB would require that the dominant saddles at positive integer~$n$ to exhibit~$\mathbb{Z}_n$ symmetry, but for the dominant saddles at small~$n$, including~$n = 0$, to break it.

Clearly the LM approach is along the lines we have in mind, as it continues the gravitational equations of motion to non-integer~$n$.  However, this continuation relies crucially on two properties.  The first is the assumption of~$\mathbb{Z}_n$ symmetry, without which it would be unclear how to express the equations of motion on a manifold with a single boundary (just as in the SK model it was unclear how to generalize the replica-symmetric ansatz~\eqref{eq:replicasymmetricq} until Parisi's breakthrough).  The second is that there is a known~$n = 1$ saddle around which the equations of motion can be perturbed to study the behavior near~$n = 1$; there is no such saddle with~$n = 0$.  These are the two primary challenges that need to be overcome in order to properly understand the role of RSB in computing gravitational free energies, and more generally any extensive quantity.

\section{Discussion}
\label{sec:disc}

We have argued that the computation of extensive quantities via a gravitational path integral should be done using a replica trick which includes contributions from connected geometries.  The inclusion of these connected saddle points dramatically changes the behavior of the theory at very low temperatures, and naturally accommodates the interpretation of  semiclassical gravity as dual to an ensemble average rather than to a particular quantum theory.  Let us now discuss open questions and natural directions for future work.

\paragraph{Ensemble Averaging in Higher Dimensions} As alluded to in Section~\ref{sec:intro}, UV corrections to the GPI may remedy the apparent lack of factorization that motivated the ensemble averaging interpretation in the first place, as discussed in the context of random matrix models and JT gravity in~\cite{SSS}.  Such a picture becomes especially crisp in higher dimensions: for example,~${\cal N}=4$ SYM is a single theory, and AdS/CFT provides numerous other examples of unitary quantum theories of gravity without the need to ensemble average.  If, however, one would like to apply the techniques of~\cite{PenShe19, AlmHar19} to higher dimensions then we must include replica wormholes, whose most obvious interpretation is of an ensemble average.  One possibility is that averaging is only genuinely necessary in certain low-dimensional theories (as was argued in e.g.~\cite{McNamara:2020uza}).  For example, the low-temperature spectrum of higher dimensional gravity (and CFTs) is perfectly well-behaved, has a unique ground state, and does not resemble a spin glass.  We do not expect to see replica wormholes or RSB dominating the free energy calculation at low temperature.  Nevertheless, it is natural to speculate that replica wormholes will contribute to $\overline{\ln Z}$ whenever we are in a regime where non-perturbative quantum gravitational corrections are important: for example, after the Page time~\cite{Pag93} or at the Hawking-Page phase transition~\cite{HawPag83}.

Another interesting possibility arises from the phenomenon of self-averaging: in a chaotic theory, the average over an ensemble of theories is often essentially harmless, as each individual instance of the ensemble is representative of the ensemble as a whole, at least for relatively coarse-grained observables.  The ensemble average in this case is interpreted as a useful calculational trick to construct a universal effective theory which governs the dynamics at low energy, but of course the UV dynamics of each individual instance of the ensemble is that of a unitary quantum theory.  Perhaps any gravitational theory which includes Euclidean wormholes should be understood as a low-energy effective theory in this sense; in this interpretation, the GPI plays the role of a convenient calculational trick for computing observables in a semiclassical limit.  Such a possibility was discussed in various forms in~\cite{PenShe19,BeldeB20,PolRoz20}.

\paragraph{Nonperturbative Completions}  At the end of Section~\ref{sec:intro}, we briefly mentioned that although a large-$N$ analysis of SYK does not exhibit a spin glass phase,~\cite{AreKhr18} showed that in a large-coupling (or low-temperature) limit that reduces to an EFT of the low-energy dynamics of SYK, saddles that correlate replicas in the computation of~$\overline{Z^m}$ become \textit{dominant} at both positive integer~$m$ as well as in the~$m \to 0$ limit, and therefore lead to a spin-glass like phase transition in this low-energy EFT.  This observation may raise a concern: if a spin glass phase can only be obtained from the SYK model by excluding the UV, is the phase transition that we have found in JT gravity eliminated by a good UV completion?  Our study of the Airy limit in Section~\ref{subsec:Airy} shows that a nonperturbative completion of JT gravity cannot eliminate the effect we have studied, since it is dominated by the universal behavior of the edge of the spectral density~$\rho(E)$.  Indeed, the recent discussion of such completions in~\cite{Joh20} explicitly finds that the two-point correlator~$\overline{Z(\beta)^{2}}$ is controlled by the contribution of connected topologies at sufficiently low temperatures, even in a nonperturbative completion.

More generally, the results of~\cite{SSS} suggest that a good nonperturbative description of JT gravity should be available in the form of a matrix model (though this completion is not unique).  Because the behavior we have studied in this paper is due to universal behavior at the spectral edge (at least at sufficiently low temperatures), we might investigate it more thoroughly by working in a toy matrix model like the Gaussian matrix integral investigated in~\cite{Oku19}.  To this end, it would be interesting to compute~$\overline{\ln Z}$ in such a model by expressing~$\ln Z = \ln \Tr e^{-\beta H}$ and then explicitly computing an average over the random matrix~$H$, without resorting to a replica trick.  We should expect to find a monotonic free energy all the way to zero temperature, with a free energy that agrees with the annealed free energy of the Airy case once the temperature becomes sufficiently (but not too) large.

\paragraph{The Emergence of Semiclassical Gravity}  A longstanding question in quantum gravity is how the (semi)classical metric~$g_{ab}$ emerges from an underlying quantum theory.  In the SK model, the partition functions~$\overline{Z^m}$ can be expressed \textit{exactly} via the introduction of the mean fields~$s_\alpha$ and~$q_{(\alpha,\gamma)}$ in~\eqref{eq:ZmSKexact}.  In a large-$N$ limit, the phase structure of the system is determined by the saddle point equations for these fields.  Importantly, they appear purely as a consequence of the disorder average; they are not fundamental in the pre-disorder theory.  (In the SYK case, the analogous fields are the auxiliary fields~$G_{\alpha\beta}(\tau_1,\tau_2)$ and~$\Sigma_{\alpha\beta}(\tau_1,\tau_2)$.)

If we are to interpret the GPI as computing a disorder average (either genuinely or in an effective description for the purpose of probing appropriately coarse-grained observables), is there a sense in which the metric should then be thought of as a mean field, with the GPI analogous to the right-hand side of~\eqref{eq:ZmSKexact}?  That is, rather than being a fundamental field of the underlying theory, is the metric a field whose existence relies fundamentally on the ensemble average?  In such a case we would interpret the ``turning on'' of connected geometries between disconnected boundaries as analogous to the ``turning on'' of the matrix~$q_{(\alpha,\gamma)}$ in the SK model.  This would give a clear meaning to the sum over topologies in the path integral, but even in the 2D models we have studied here it is unclear how this interpretation would incorporate a UV completion.

\paragraph{RSB and the Parisi Ansatz in Gravity} In the 2D models studied in this paper, the need for replica wormholes in the free energy (or more generally, any extensive quantity) is clear, and we have discovered hints of RSB.  These suggest that gravity has some features analogous to a glassy phase just at the edge of semiclassicality.  Since the gravitational path integral is in general -- and in this regime in particular -- of clear interest, clearly one important extension of our analysis would be the construction of a gravitational analogue of the Parisi ansatz for RSB.  Of course, because we did not work in any saddle point approximation, we did not consider classical equations of motion.  The resulting lack of any saddles to analyze for stability or to continue to~$m = 0$ makes it difficult to explore the structure of RSB in any detail.  In particular, the fact that (pure) JT gravity replica wormholes do not exist as solutions to any classical equations of motion suggests that there may be no way to study RSB in JT gravity in a way analogous to conventional spin glass systems (though admittedly the possibility of a phase transition at~$m < 1$ means that the lack of on-shell wormholes for integer~$m$ does not necessarily exclude on-shell analytically continued wormholes for~$m$ near zero).  A natural question, then, is whether there exist models of gravity that are sufficiently simple to allow for the continuation of classical equations of motion to~$m = 0$, but sufficiently complex to still exhibit a phase transition.  In other words, it would be valuable to find a gravitational model in which the effects of Euclidean wormholes can be disentangled from those of disconnected geometries with higher genus (analogous to the case of~$\widehat{\mathrm{CGHS}}$, in which higher genera don't appear at all).

In such a model, we might imagine that the correct ``gravitational'' Parisi ansatz is a multi-branched wormhole connecting the various disconnected boundaries with wormholes of different sizes, with these sizes left as variational parameters with respect to which the free energy should be extremized.  In the case of a near-extremal black hole (and consequently low temperature), the picture might be reminiscent of AdS fragmentation~\cite{MalMic98}, in which the AdS$_2$ throat can fragment into many throats or disconnected universes.  Understanding how this story works in gravity would be especially illuminating because the Parisi function~$q(x)$, which plays the role of an order parameter for the spin glass phase transition in the SK model, also probes the structure of microstates of the model.  An analogous function in gravity could shed light onto the details of the underlying (that is, pre-disorder-average) theory.

\section*{Acknowledgements}

It is a pleasure to thank D.~Harlow and J. Sully for helpful discussions and D.~Anninos and D.~Stanford for useful comments on an early version of this paper. The work of NE is supported by the Office of High Energy Physics of U.S. Department of Energy under grant Contract Number DE-SC0012567 and by the MIT department of physics. 
Research of SF and AM is supported in part by the Simons Foundation Grant No.~385602 and the Natural Sciences and Engineering Research Council of Canada (NSERC), funding reference number SAPIN/00032-2015.

\appendix

\section{Airy Limit}
\label{app:Airy}

In order to make this paper more self-contained, in this Appendix we briefly review the relevant results on the low-temperature limit of JT gravity discussed in Section~\ref{subsec:Airy}.

The starting point is Mirzakhani's formula for the Weil-Peterson volumes~$V_{g,m}$ appearing in~\eqref{subeq:ZgmVgm}~\cite{Mir06b}:
\bea
V_{g,m}(\{b_i\}) &= \frac{1}{(3g-3+m)!} \int_{\overline{\mathcal{M}}_{g,m}} \left(2\pi^2 \kappa + \frac{1}{2} \sum_{i = 1}^m b_i^2 \psi_i\right)^{3g-3+m}, 
\label{Vgm}
\\
	&= \sum_{\mathclap{\substack{\bm{\alpha},p \\ |\bm{\alpha}| + p = 3g-3+m}}} \frac{(2\pi^2)^p}{2^{|\bm{\alpha}|} \alpha_1! \cdots \alpha_m! p!} \, b_1^{2\alpha_1} \cdots b_m^{2\alpha_m} \int_{\overline{\mathcal{M}}_{g,m}} \psi_1^{\alpha_1} \cdots \psi_m^{\alpha_m} \kappa^p,
\eea
where~$\overline{\mathcal{M}}_{g,m}$ is the Deligne-Mumford compactification of the moduli space of constant-negative curvature Riemann surfaces of genus~$g$ with~$m$ geodesic boundaries of lengths~$b_i$, the~$\psi_i$ are Chern classes,~$\kappa$ is the first Mumford-Morita-Miller class on~$\overline{\mathcal{M}}_{g,m}$, and we use the notation~$\bm{\alpha} = \{\alpha_1, \ldots, \alpha_m\}$ and~$|\bm{\alpha}| = \sum_{i = 1}^m \alpha_i$; see e.g.~\cite{Do11} for a review.  The quantity in parenthesis in equation (\ref{Vgm}) is the Weil-Peterson symplectic form on the moduli space of bordered Riemann surfaces.
Because~$V_{g,m}(\{b_i\})$ is a polynomial in the~$b_i$, when inserted into~\eqref{subeq:ZgmVgm} we may explicitly perform the integrations over the~$b_i$ to obtain
\be
Z_{g,m}(\beta) = \sum_{\mathclap{\substack{\bm{\alpha},p \\ |\bm{\alpha}| + p = 3g-3+m}}} \frac{(2\pi^2)^p}{p! (2\pi)^{m/2}} \, \beta^{3g-3 + 3m/2-p} \int_{\overline{\mathcal{M}}_{g,m}} \psi_1^{\alpha_1} \cdots \psi_m^{\alpha_m} \kappa^p.
\ee
At low temperatures, the leading-order behavior of~$Z_{g,m}$ comes from the terms in the sum with~$p = 0$; keeping only these terms,~\eqref{eq:JTgenusexpansion} gives
\be
\label{eq:PmAirypartial}
\Pcal_{\mathrm{conn},m}(\beta) = \left(\frac{\beta e^{-2S_0/3}}{2\pi}\right)^{m/2} \sum_{g = 0}^\infty \sum_{\substack{\bm{\alpha} \\ |\bm{\alpha}| = 3g-3+m}} \left(\beta e^{-2S_0/3}\right)^{3g-3 + m} \int_{\overline{\mathcal{M}}_{g,m}} \psi_1^{\alpha_1} \cdots \psi_m^{\alpha_m} + \cdots,
\ee
where the ellipses denote terms that are subleading at low temperature.

The sum over genus was computed in~\cite{Oko01}.  To express it, introduce the function
\begin{multline}
\Ecal^{(m)}(x_1, \ldots, x_m) \equiv \frac{\exp\left(\sum_{i = 1}^m x_i^3/12\right)}{(4\pi)^{m/2} \sqrt{\prod_{i = 1}^m x_i}} \\ \times \int_{s_i \geq 0} d^m s \, \exp\left(-\sum_{i=1}^m \frac{(s_i-s_{i+1})^2}{4x_i} - \frac{1}{2} \sum_{i = 1}^m (s_i + s_{i+1}) x_i \right),
\end{multline}
where we identify~$s_{m+1} \equiv s_1$.  By construction~$\Ecal^{(m)}(\{x_i\})$ is invariant under cyclic reorderings of the~$x_i$; let us therefore define the function
\be
\Ecal^{(m)}_\mathrm{sym}(x_1, \ldots, x_m) = \frac{1}{m} \sum_{\sigma \in \mathbb{S}_m} \Ecal^{(m)}(x_{\sigma(1)}, \ldots, x_{\sigma(m)}),
\ee
which by construction is invariant under any permutation of the~$x_i$ (here~$\mathbb{S}_m$ is the permutation group of order~$m$).  Next, let~$\Pi_m$ be the set of all partitions of~$\{1, \ldots, m\}$ into disjoint unions of subsets, for any~$q \in \Pi_m$ let~$\ell(q)$ be the number of blocks in~$q$, and let~$x_q$ be the set of size~$\ell(q)$ formed by summing the~$x_i$ over the blocks of~$q$.  For example,
\be
\Pi_3 = \{ \{1,2,3\}, \{1,2\} \sqcup \{3\}, \{1,3\} \sqcup \{2\}, \{2,3\} \sqcup \{1\}, \{1\} \sqcup \{2\} \sqcup \{3\}\},
\ee
and if~$q = \{1,2\} \sqcup \{3\} \in \Pi_3$,~$\ell(q) = 2$ and~$x_q = \{x_1 + x_2, x_3\}$.  Using this notation, we now define
\be
G^{(m)}(x_1, \ldots, x_m) \equiv \sum_{q \in \Pi_m} (-1)^{\ell(q)+1} \Ecal^{(\ell(q))}_\mathrm{sym}(x_q),
\ee
so for instance
\begin{subequations}
\be
G^{(2)}(x_1,x_2) = \Ecal^{(1)}_\mathrm{sym}(x_1 + x_2) - \Ecal^{(2)}_\mathrm{sym}(x_1,x_2),
\ee
\begin{multline}
G^{(3)}(x_1,x_2,x_3) = \Ecal^{(1)}_\mathrm{sym}(x_1 + x_2 + x_3) - \Ecal^{(2)}_\mathrm{sym}(x_1 + x_2, x_3) - \Ecal^{(2)}_\mathrm{sym}(x_1 + x_3, x_2) \\ - \Ecal^{(2)}_\mathrm{sym}(x_2 + x_3, x_1) + \Ecal^{(3)}_\mathrm{sym}(x_1, x_2, x_3).
\end{multline}
\end{subequations}
The main result of~\cite{Oko01} can then be expressed as
\be
\sum_{g = 0}^\infty \sum_{\substack{\bm{\alpha} \\ |\bm{\alpha}| = 3g-3+m}} x_1^{\alpha_1} \cdots x_m^{\alpha_m} \int_{\overline{\mathcal{M}}_{g,m}} \psi_1^{\alpha_1} \cdots \psi_m^{\alpha_m} = \frac{(2\pi)^{m/2}}{\sqrt{\prod_{i = 1}^m x_i }} \, G^{(m)}\left(\frac{x_1}{2^{1/3}}, \ldots, \frac{x_m}{2^{1/3}}\right).
\ee
Applying this result to~\eqref{eq:PmAirypartial} with~$x_i = \beta e^{-2S_0/3}$ for all~$i$, we thus obtain
\be
\label{eq:PmAiry}
\Pcal_{\mathrm{conn},m}(\beta) = G^{(m)}\left(\frac{\beta e^{-2S_0/3}}{2^{1/3}}, \ldots, \frac{\beta e^{-2S_0/3}}{2^{1/3}}\right) + \cdots.
\ee
The low-temperature subleading corrections to~\eqref{eq:PmAiry} were computed for the~$m = 1$ case in~\cite{OkuSak19}, and are expressed schematically in~\eqref{eq:JTlowtemp}; the~$m = 2$ corrections were computed in~\cite{OkuSak20}.

\bibliographystyle{jhep}
\bibliography{all}

\end{document}